\documentclass[manuscript,screen]{acmart}

\AtBeginDocument{%
  \providecommand\BibTeX{{%
    \normalfont B\kern-0.5em{\scshape i\kern-0.25em b}\kern-0.8em\TeX}}}

\setcopyright{acmcopyright}
\copyrightyear{2026}
\acmYear{2026}
\acmDOI{XXXXXXX.XXXXXXX}

\acmSubmissionID{2897}

\graphicspath{ {Images/} }
\usepackage{enumerate}
\usepackage{amsmath}
\usepackage{gensymb}
\usepackage{subcaption}
\usepackage{tabularx}
\usepackage{multicol}
\usepackage{multirow}
\usepackage{graphicx}
\usepackage{longtable}
\usepackage{lipsum}
\usepackage{pdfpages}
\usepackage{ulem}
\usepackage{censor}

\definecolor{emred}{RGB}{250, 0, 0}
\definecolor{emgreen}{RGB}{0, 180, 0}

\begin{document}
\title{Auxilio and Beyond: Comparative Study and Design Guidelines for Head Movement-based Assistive Mouse Controllers}

\author{Mohammad Ridwan Kabir}
\email{ridwan.kabir@queensu.ca}
\orcid{0000-0002-9631-1836}
\authornotemark[1]
\affiliation{%
  \institution{~iStudio Lab,~School of Computing,~Queen's University}
  \city{Kingston}
  \state{Ontario}
  \country{Canada}
  \postcode{K7L 3N6}
}
\author{Mohammad Ishrak Abedin}
\authornote{Both authors contributed equally to this research.}
\email{ishrakabedin@iut-dhaka.edu}
\orcid{0000-0002-7825-0339}

\author{Rizvi Ahmed}
\email{rizviahmed@iut-dhaka.edu}
\orcid{0000-0002-2089-5994}

\author{Saad Bin Ashraf}
\email{saadashraf@iut-dhaka.edu}
\orcid{0000-0002-1455-9038}
\affiliation{%
  \institution{~Network and Data Analysis Group (NDAG),~Department of Computer Science and Engineering (CSE),~Islamic University of Technology (IUT)}
  \city{Board Bazar}
  \state{Gazipur}
  \country{Bangladesh}
  \postcode{1704}
}

\author{Hasan Mahmud}
\email{hasan@iut-dhaka.edu}
\orcid{0000-0003-4375-6943}

\author{Md. Kamrul Hasan}
\email{hasank@iut-dhaka.edu}
\orcid{0000-0003-1295-7945}

\affiliation{%
  \institution{~Systems and Software Lab (SSL),~Department of Computer Science and Engineering (CSE),~Islamic University of Technology (IUT)}
  \city{Board Bazar}
  \state{Gazipur}
  \country{Bangladesh}
  \postcode{1704}
}
\renewcommand{\shortauthors}{Kabir, et al.}

\begin{abstract}
    Upper limb disability due to neurological disorders or other factors restricts computer interaction for affected individuals using a generic optical mouse. This work reports the findings of a comparative evaluation of Auxilio, a sensor-based wireless head-mounted Assistive Mouse Controller (AMC), that facilitates computer interaction for such individuals. Combining commercially available, low-cost motion and infrared sensors, Auxilio utilizes head movements and cheek muscle twitches for mouse control. Its performance in pointing tasks with subjects without motor impairments has been juxtaposed against a commercially available and patented vision-based head-tracking AMC developed for similar stakeholders. Furthermore, our study evaluates the usability of Auxilio using the System Usability Scale, supplemented by a qualitative analysis of participant interview transcripts to identify the strengths and weaknesses of both AMCs. Experimental results demonstrate the feasibility and effectiveness of Auxilio, and we summarize our key findings into design guidelines for the development of similar future AMCs.
\end{abstract}

\ccsdesc[500]{Human-centered computing~Pointing devices}
\ccsdesc[500]{Human-centered computing~Accessibility technologies}
\ccsdesc[500]{General and reference~Experimentation}
\ccsdesc[500]{General and reference~Evaluation}
\ccsdesc[500]{General and reference~Performance}

\ccsdesc[500]{Human-centered computing~Accessibility technologies}
\ccsdesc[300]{Human-centered computing~Pointing}
\ccsdesc[300]{Human-centered computing~Usability testing}

\keywords{assistive technology, assistive mouse controller, upper limb disability, wearable sensors, pointing device}


\maketitle

\section{Introduction}    
    Upper limb disability is the complete or partial motor dysfunction of the upper limb, potentially caused due to --- stroke \cite{c2,c3,c4,c5}, amputation \cite{c16,c17,c18}, etc. The under-utilized residual capabilities of the disabled upper limb might hamper the lives of those affected in terms of both \textit{activity limitations} and \textit{participation restrictions} \cite{c19}. Research has shown that, unlike a normal person, these residual sensory abilities of disabled individuals intensify over time, compensating for their lost ability \cite{c28, c29, c30, c31, c32}. Eventually, they learn to utilize these abilities to accomplish different tasks in their daily lives \cite{c33, c34, c35, c36}. Although physically capable individuals can seamlessly use generic handheld pointing devices (e.g. an optical mouse) for interacting with a computer, people with upper limb disability require Assistive Mouse Controller (AMC) as an alternative input modality for the same \cite{kabir2023acceptability}. 
    
    These AMCs can either be vision-based \cite{c39, c40, c41, c42, SmyleMouse1, SmyleMouse2, SmyleMouse3}, Electromyography (EMG)-based \cite{c43, c44, c45, c46}, Electrooculogram (EOG)-based \cite{c24, c27, c47, c48}, Wearables Sensor-based \cite{c45, c49, c50, c51}, or a combination of these \cite{c83, c84}. The user interactions with these AMCs are carefully designed to allow users to utilize their residual motor capabilities with minimal effort. For example, patients with  Amyotrophic Lateral Sclerosis (ALS) suffer from motor dysfunctionality of their body except for the eye muscle movement, which remains unaffected throughout the disease \cite{c12, c24, c27, nijssen2017motor}. On the contrary, the motor capabilities of the head and neck are retained in upper limb amputees. Therefore, people with ALS are unable to use AMCs that utilize head-motion as an input modality. However, they can seamlessly use vision-based, or EOG-based ones \cite{c12, c24, c27, nijssen2017motor} with effort commensurate with their residual capabilities.
    
    While the existing AMCs try to cater to different needs, they face limitations such as but not limited to --- sensitivity to lighting conditions \cite{sabab2022vis, c50}, frequent calibration needs \cite{c88}, noise interference \cite{c87}, hygiene concerns \cite{c100}, and unreliable performance in fluctuating environments \cite{c24, c87}, highlighting the need for more user-friendly and robust alternatives.
    Hence, the design, development, and evaluation of alternative input modalities for individuals with upper limb disability is still of interest to the research community \cite{c19}. Along with that, the importance of analyzing the comparative performance, usability, and feedback of such an AMC from the users' point of view cannot be overemphasized \cite{c133}. 

    This paper presents a significant iteration in the development of an assistive device, Auxilio \cite{kabir2023acceptability}, a prototype of a sensor-based head-mounted wireless AMC purpose built for people with intact motor capabilities of their --- (1) head and neck muscles facilitating head movements, and (2) facial muscles facilitating cheek twitches or smile gestures \cite{c102}, as our primary users. We have kept our focus narrow to a selected group as the broad spectrum of upper-limb disability would require specialized solutions based on the specific residual capabilities and restrictions. Auxilio combines a low-cost Commercial Off-The-Shelf (COTS) Inertial Measurement Unit (IMU) for controlling the mouse cursor with absolute head movements and infrared sensors for actuating mouse clicks with cheek muscle twitches.  We compare its performance against Smyle Mouse \cite{SmyleMouse1, SmyleMouse2, SmyleMouse3}, a commercially available and patented camera-based AMC that controls the cursor with tracked head movements and registers clicks with smile gestures. We acknowledge that comparing Auxilio against another sensor-based AMC would have been ideal; however, we were unable to procure any such AMC in our locality. Hence, we compared it against the Smyle Mouse, which is similar in input modality and is a recognized, patented system.

    While our initial plan involved developing Auxilio in a functionality-centered way and testing it against first-hand stakeholders, practical field testing failed to provide insightful information due to the stakeholders' lack of domain knowledge. Additionally, getting continuous access to individuals with upper limb disability is economically unfeasible in the context of our country. Consequently, we decided to proceed with a refined iterative approach and updated design rationale (see \autoref{subsec:des_ration}), carrying out our technical investigation involving expert proxies (without motor impairments) with adequate domain knowledge to provide actionable feedback. Since our target demographic has fully-retained motor control over their head-neck and facial muscles, able-bodied domain experts were considered vital for distinguishing between \textit{system error} and \textit{user error} while preserving physiological validity of the study (experts and primary users rely on the exact same intact muscle groups \cite{kikkert2016neural, makin2013plasticity}). Since Auxilio is fairly new to the domain, we believe its performance, usability, and design limitations need to be analyzed further before it can be made available to its primary stakeholders.

    In Human-Computer Interaction (HCI), a pointing task is considered as the user interaction for selecting any element on a user interface with any pointing device (e.g. an optical mouse) \cite{kabir2022antasid}. User performance with AMCs have been previously evaluated with pointing tasks while applying Fitt's Law \cite{kabir2022antasid, c44, c50}. To this end, we conducted a within-subject \textit{Point and Click} experiment involving $10$ domain-experts featuring a balloon-popping game, \textit{Popper} \cite{kabir2022antasid}. We followed this up with a qualitative analysis of the interview transcripts to find strengths, weaknesses, and future directives. We further investigated the usability of Auxilio in comparison with Smyle Mouse, leveraging the System Usability Scale (SUS), a $10$-item $5$-point Likert scale-based closed-questionnaire \cite{c60, c61, c62, c63, c64, c65}. We firmly believe the future directives to be valuable to anyone trying to create similar AMCs in the future and consider them the highlight of this work. To the best of our knowledge, this work is the first of its kind in the literature. In summary, the main contributions of this study in the domain of accessibility to computer interaction for the physically disabled community are as follows:
    \begin{enumerate}[(a)]
        \item We detailed the design rationale and hardware-software implementation of Auxilio, a sensor-based head-mounted AMC tailored for individuals with upper limb disabilities  but retained head motor control. 
        \item We evaluated the feasibility and performance of Auxilio in comparison to the Smyle Mouse, a patented, camera-based head-tracking AMC, through controlled pointing tasks.
        \item We conducted a comparative usability analysis of both AMC technologies using the System Usability Scale (SUS) to highlight the usability challenges and advantages of each.
        \item We carried out a qualitative analysis of user feedback to identify key strengths and limitations of the two AMC types, providing recommendations and distilling design guidelines for developing similar AMCs.
    \end{enumerate}
   
\section{Related Works} 
    The recent technological advancements have shaped the design, and development of Assistive Mouse Controllers (AMCs) for physically challenged individuals into a prominent research area \cite{c19}. Interaction data from such individuals may be recorded either using computer vision, Electromyography (EMG), Electrooculogram (EOG), or wearable sensors. In this section, we elaborate on the existing state-of-the-art AMCs and justify our scope and motivation behind this study.
        
    \subsection{Vision-based AMCs: Design Tradeoffs and Limitations}
        Vision-based AMCs map facial or eye gaze features, recorded using eye trackers or webcams, to screen coordinates for cursor control \cite{c75, c76, c77, SmyleMouse1, SmyleMouse2, SmyleMouse3, sabab2022vis} after proper calibration. Common click actuation methods include dwell-time mechanisms or gestures like \textit{eye wink}, \textit{blink}, or \textit{smile}. For instance, Smyle Mouse tracks the user's nose for cursor movement and uses smile gestures for left-click events, preceded by a gesture-based calibration phase, while offering a dwell-time-based click actuation alternative \cite{SmyleMouse1, SmyleMouse2, SmyleMouse3}. However, events like left-click, right-click, double-click, drag, etc., can also be mapped with the gesture from a separate UI. Zhang et al. \cite{c39} used an eye tracker for gaze-based cursor control with dwell-time clicking via a virtual UI for Mouse/Keyboard simulation. Other approaches have utilized optical mouse sensors for gaze tracking instead of dedicated cameras \cite{c40, c41}.Although common, in practice \textit{dwell-time}-based click actuation suffers from unwanted actuation of mouse clicks due to eye gaze fixation, generally known as the \textit{Midas Touch} problem \cite{c78}. To mitigate this, alternatives like muscle or eyebrow shrugging have been implemented in Camera Mouse \cite{c80, c81} and ClickerAid \cite{c82}. Rajanna et al. \cite{c83, c84} developed a hybrid system using eye gaze for pointing and a pressure-sensitive footwear for clicking, designed for users with upper-limb impairments.
        
    \subsection{Electromyography (EMG)-based AMCs: Signal Complexity and User Burden}
        Electromyography (EMG) signals refer to the measurement of very low electric potentials generated due to muscle contractions with electrodes placed noninvasively on the skin, where the signal amplitudes are proportional to the exerted muscle force \cite{c85}. For people with upper limb disabilities, EMG signals can be retrieved from the contraction of the residual muscles \cite{c86, c44}, which can then be used to determine the type of intended motion. Researchers have explored the feasibility and performance of a myoelectric cursor control for amputees with the help of a myoelectric armband  \cite{c43}. %
        For making computers accessible to people with high-level spinal cord injury, researchers in \cite{c44} have proposed two cursor control algorithms using a single-site surface EMG sensor and evaluated performance through a pointing task-based experiment utilizing Fitts’ law.
        
        
    
    \subsection{Electrooculogram (EOG)-based AMCs: Low Resolution and Ergonomic Concerns}
        An Electrooculogram (EOG) is used to measure the corneal-retinal Transepithelial Potential Difference (TEPD) with the help of noninvasive electrodes placed around the human eyes \cite{c26, c47, c48}, which can be utilized to implement mouse control and clicking mechanisms.. 
        Several research have been carried out with EOG-based AMCs to facilitate interaction with a computer for people suffering from neurodegenerative disorders \cite{c95, c96, c97, c98, c99}. To facilitate typing through recognition of eye movement patterns, real-time EOG-based systems have been developed \cite{c24, c27, c89, c90, c91, c92} for people with ALS \cite{c25, c26}.%
        

    \subsection{Sensor-based Wearable AMCs: Towards Non-Intrusive Design}
        Sensor-based wearable AMC technologies leverage residual motor functionalities as alternative input modalities, among which head movement is a natural, effective, and the most common one for moving a cursor \cite{kabir2023acceptability, c45, c49, c50, c51, qin2024customized}. Other options include utilizing tongue muscle movement \cite{c100}, and Brain Computer Interfaces (BCI) \cite{c101}. Velasco et al. \cite{c51} developed an AMC for people with cerebral palsy, where the subjects can move the cursor through their head movements and a pointing facilitation algorithm, \textit{MouseField}. Gorji et al. \cite{c50} used Infrared (IR) sensors mounted inside a collar, which is wearable on the user’s neck below the chin level, for measuring the user’s range of head tilt motion. 
        
        For mouse click actuation using such AMCs, diverse approaches such as --- \textit{dwell-time}-based \cite{c51}, EMG-based \cite{c45}, BCI-based \cite{c49, c101}, flex sensors activated using cheek muscle twitches \cite{c102}, using tongue muscles to click a joystick \cite{c100}, etc., have also been explored in the literature. Yamamoto et al. \cite{c103} further explored a wearable strain sensor as a click actuation device through a single case study involving a user with mixed types of cerebral palsy. Performance for cursor control is often evaluated using path-following tasks and Fitts’s law \cite{c50}.

        \begin{figure}[htbp]
            \centering
            \includegraphics[width=.75\textwidth]{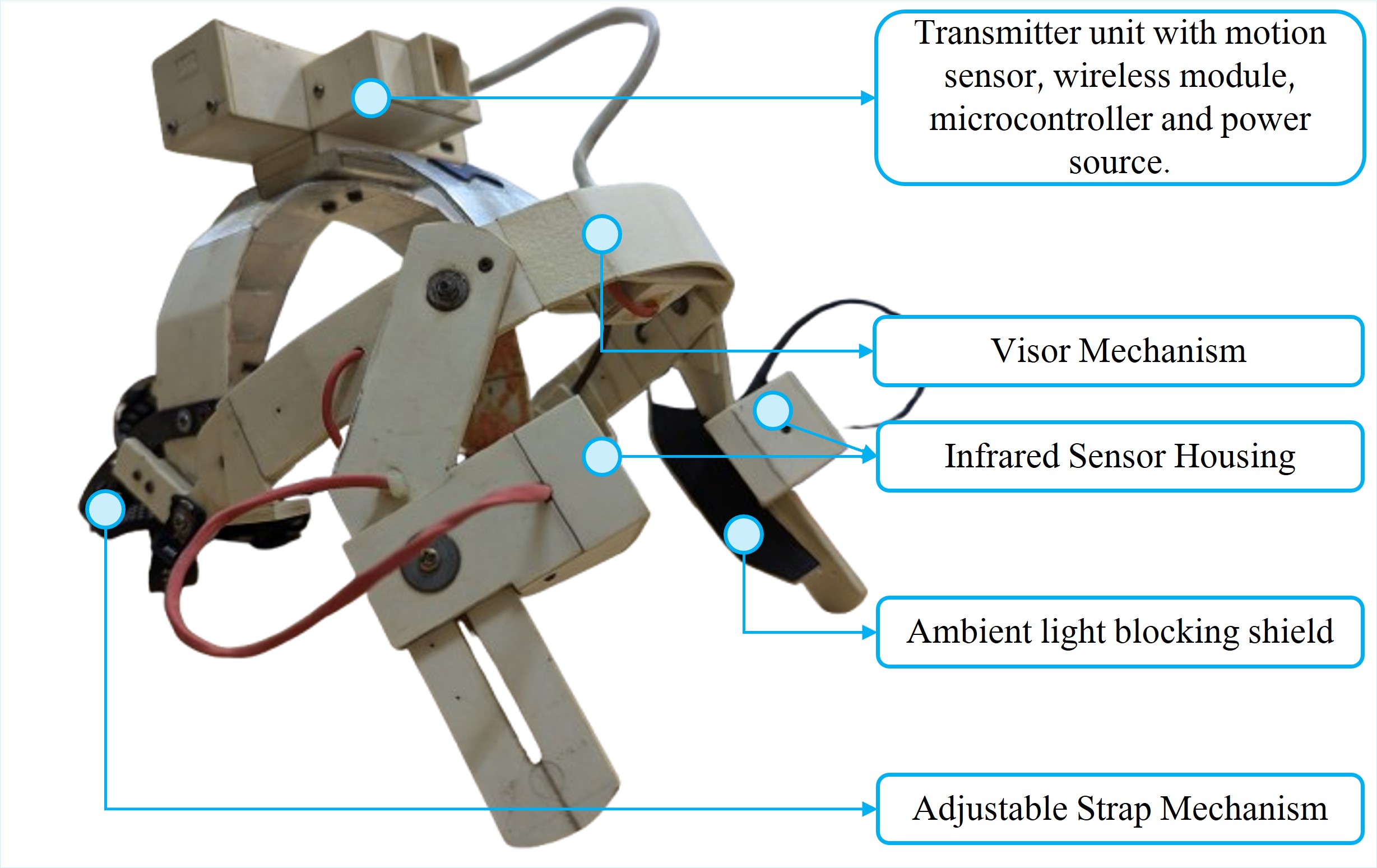}
            \caption{Prototype of Auxilio, a sensor-based head-mounted wireless Assistive Mouse Controller (AMC) that uses absolute head movements for mouse cursor control and cheek muscle twitches for mouse click actuation using an Inertial Measurement Unit (IMU) and infrared sensors.}
            \Description{A labeled image depicting the constituent elements of Auxilio, such as the transmitter unit with motion sensor, wireless module, microcontroller, and power source; the visor mechanism, infrared sensor housing, ambient light blocking shield, and adjustable strap mechanism.}
            \label{fig:Auxilio Full}
        \end{figure}
        

    \subsection{Current Research Gaps: Identifying the Design Space of a Novel Wearable AMC}
        To identify the potential design space for a novel wearable AMC, we need to first understand the limitations of the current state of the art technologies as discussed in the previous sections.
        
        For vision-based AMCs, proper lighting is critical for calibrating and accurately detecting facial \cite{c42, c75, c76, c77} or eye gaze \cite{c39, c40, c41} features. Gaze tracking presents further challenges --- image resolution, variable lighting, dependence on glasses, and users' skin complexion. Additionally, human eyes are not the most accurate pointing device \cite{c58} and can only fixate on a single point at a given time. Coordinating gaze-based cursor movement and click actuation (e.g., during drag-and-drop) is also difficult. As stated earlier, \textit{dwell-time}-based click actuation introduces the \textit{Midas Touch} problem \cite{c78}, causing unwanted selection. Finally, limited working range of eye trackers and webcams restricts user movement, requiring a fixed distance from the workstation.
        
        EMG signals, however, are susceptible to noise and depend heavily on precise electrode placement \cite{c87} for accurate gesture recognition. A further limitation is that the signals are typically weak and exhibit significant inter-user variability \cite{c88}, necessitating user-specific calibration and model training for each new operator. While deep learning techniques \cite{c46} have been proposed to mitigate this need, the associated computational expense appears unreasonable for the core objective of an AMC.
        
        Although EOG is a promising, low-cost AMC technology under active research, it is limited by low spatial resolution, as estimating absolute gaze position is complicated by bio-potential noise \cite{c27, c93, c95}. Signal characteristics also vary widely due to differences in electrode number, type (dry/wet), material, and placement \cite{c24, c90, c91, c93, c94, c95}, adding significant pre-processing complexity \cite{c93}. Finally, continuous eye movements for cursor control may pose certain health issues, thereby, affecting the user’s performance and comfort.

        Existing sensor-based wearable AMCs also present limitations. For instance, the device in \cite{c100} is placed inside a user’s mouth through a retainer, raising health and hygiene issues and risks tongue muscles fatigue from prolonged usage. Its performance is also questionable due to unrealistic IR sensor placement \cite{c50}, which makes readings susceptible to lighting fluctuations. Furthermore, in scenarios that require rapid movements of the mouse cursor, their device may not be a viable option. Similarly, the device developed by Yamamoto et al. \cite{c103} can not be used by individuals with amputated or disabled upper limbs.
    
        In light of the above discussions, a new sensor-based AMC can address several limitations in existing technologies. Vision-based AMCs often face challenges with lighting conditions, calibration, and the ``Midas Touch'' problem in dwell-time-based click mechanisms. Similarly, EMG and EOG-based systems suffer from external noise, complex calibration, and low spatial resolution. Wearable sensor-based AMCs, while promising, often require intrusive setups or cumbersome calibration processes. A non-intrusive, head-motion-controlled AMC using IR sensors for clicks can offer a more intuitive, less invasive, and less intrusive solution, improving usability and reducing calibration complexity.
                
\section{Designing and Developing AUXILIO}
    With an aim to resolve some, if not all, of the limitations of the existing AMC technologies in the perspective of our target demographic, we designed and developed a fully functional prototype of Auxilio (\autoref{fig:Auxilio Full}), a head-mounted AMC that --- (1) facilitates mouse cursor control using head motions registered with an IMU and (2) actuates mouse clicks with cheek muscle twitches detected using infrared sensors. The system architecture comprises three core components: a transmitter unit, a receiver unit, and device driver software, each with distinct responsibilities yet tightly integrated through a well-defined data processing workflow. Auxilio was designed with three core design rationale in mind, which we emerged from our iterative design process (\autoref{fig:overall_design_workflow}) --- \textbf{DR1:} comfort-driven form factor, \textbf{DR2:} intuitive interaction mechanisms, and \textbf{DR3:} deterministic system behavior. In this section, we elaborate how these design principles led to the specific design choices and workflow for Auxilio.
    
    \subsection{Design Rationale} \label{subsec:des_ration}
        \subsubsection{DR1: Comfort-Driven Form Factor}
            The transmitter unit (\autoref{fig:Auxilio Full}), worn on the user’s head like a helmet, houses the necessary electronics (microcontroller, wireless communication module, battery, the Inertial Measurement Unit (IMU), infrared sensors, etc.). Since the shape of the human head and fluffiness of cheeks vary across individuals \cite{kabir2023acceptability}, Auxilio has adjustable head-straps to ensure it fits across different head-shapes. The IMU, based on the MPU9250 sensor, captures real-time head movements in terms of yaw and pitch within ergonomic limits. Since the infrared sensors detect cheek muscle twitches, the received signal will vary for different cheek shapes. Thus they are placed inside a sensor housing that can be moved up and down on respective rails on both cheeks to record signals properly. A culmination of these design choices ensure a comfort-driven form factor for Auxilio.  
    
        \subsubsection{DR2: Intuitive Interaction Mechanisms}
            The IMU-registered yaw and pitch movements of the head are utilized to move the cursor left/right and up/down, respectively, while the infrared sensors detect subtle cheek muscle contractions to actuate mouse click events in real time. The underlying software allows these intuitive interactions to be performed simultaneously, enabling full mouse functionality. An added layer of control is implemented via gesture-based toggling mechanisms. The gesture controller allows users to temporarily enable or disable the mouse interaction mode. For instance, by nodding downward and activating both cheek sensors, the mouse can be disabled to avoid unintentional movements during rest. Re-enabling is performed via an upward nod in conjunction with bilateral cheek activation. 
    
        \subsubsection{DR3: Deterministic System Behavior}
            Auxilio undergoes a calibration step before each session, lasting a few seconds, to record the user's neutral head orientation and infrared sensor readings as a baseline reference, ensuring consistent motion mapping throughout the session. Cheek-based click detection is handled by comparing the rate of changes of infrared sensor values against predefined change rate thresholds, making it lighting condition and change independent. Release events are triggered when the respective values fall below the threshold, emulating standard mouse button release behavior. Sensor readings are wirelessly transmitted to the receiver unit through the widely used NRF24L01+ 2.4GHz radio communication module \cite{semiconductor2008nrf24l01+}, connected to a host computer, mapping them to system-level calls for controlling the mouse via a driver. Together, these features create a predictable system that forms a reliable digital response for every intentional physical gesture.
    
    \subsection{Device Workflow}
\begin{figure}[htbp]
            \centering
            \includegraphics[width=0.5\linewidth]{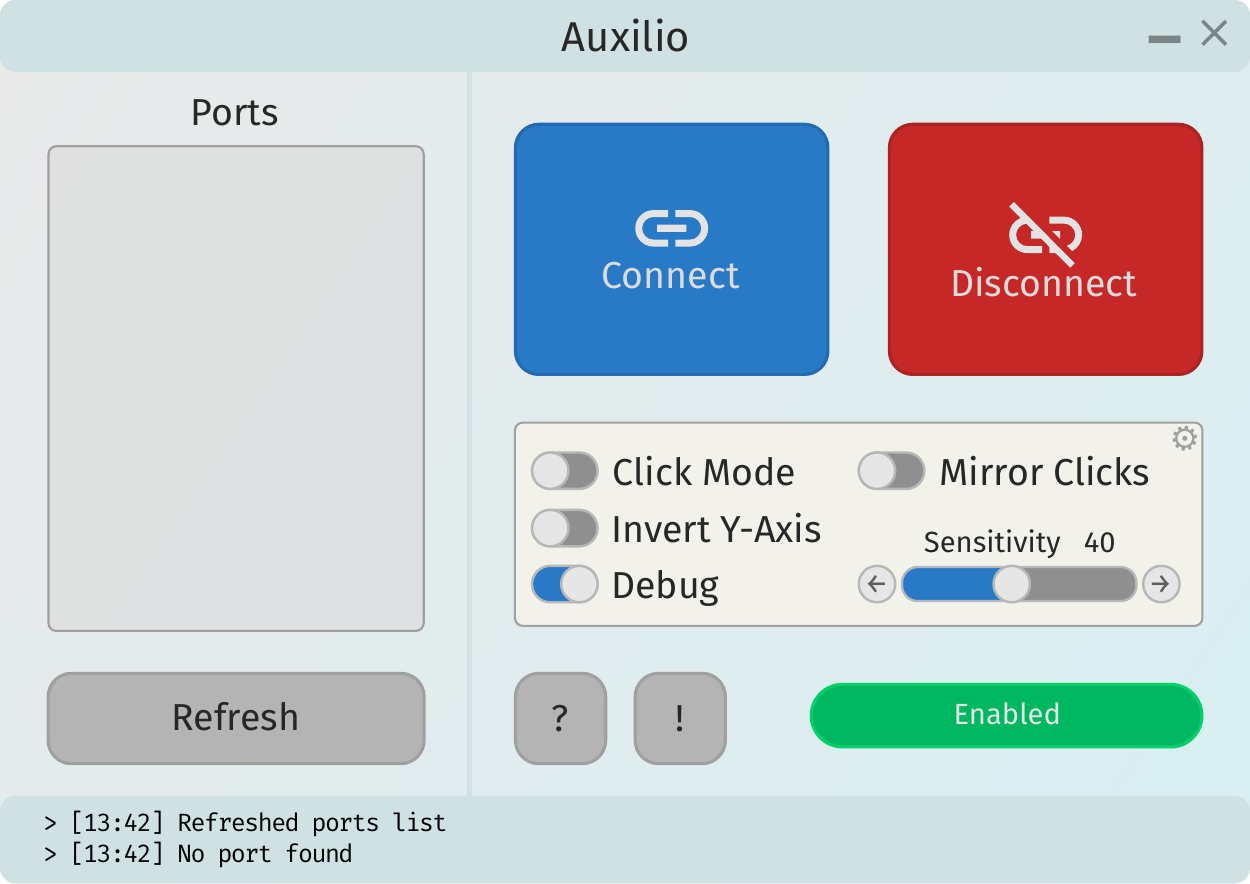}
            \caption{User Interface of the driver}
            \Description{An image of the the user interface of Auxilio. It contains a list of available ports to connect to in the left side for multi-device testing. Right side contains large rounded corner buttons for connection and disconnection of the device. Below the button, there is a settings panel with different configuration options. Below that sits buttons for about and help, along with an indicator to show whether the device is connected or disconnected. The bottom pane of the UI contains an output section to deliver status messages to the users.}
            \label{fig:auxilio_ui}
        \end{figure}

\begin{figure}[htbp]
            \centering
            \includegraphics[width=.8\textwidth]{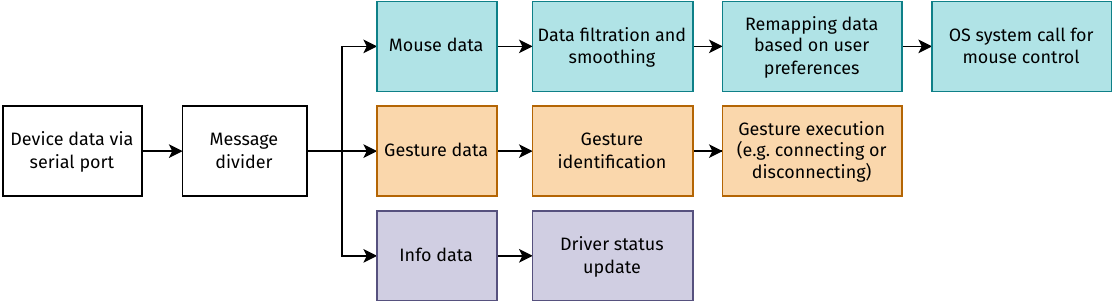}
            \caption{A workflow diagram of the driver for Auxilio.}
            \Description{A flow diagram of the workflow of Auxilio driver. The received data from the serial port goes through a message divider, dividing them into mouse, gesture, and info data. Mouse data goes through filtering, smoothing, remapping before getting transformed into OS call for mouse movment. Gesture data goes through an identification process to classify specific gesture and then it is executed. Info data is used to update driver status.}
            \label{fig:driver_workflow}
        \end{figure}

        The device driver for Auxilio serves as a bridge between the wirelessly-received data from the device and the movement of the mouse cursor by the operating system, while allowing the users to control specific behaviors of the device and facilitating basic gesture recognition. It contains a graphical user interface developed using C\# in .NET (\autoref{fig:auxilio_ui}) and is currently tested in Windows. The user interface includes controls for \textit{connecting/disconnecting} the device and tuning mouse movement and clicks based on user preferences. During the operation phase, the driver receives the device data via the serial port where the receiver unit is connected, which are treated as message strings in a pre-specified format. Upon receiving a message, a divider is responsible for separating out the message into one of the three categories, \textit{mouse data}, \textit{gesture data}, or \textit{info data}. The workflow of the driver is depicted in \autoref{fig:driver_workflow}.

        \subsubsection{Mouse Data}
            A \textit{mouse data} message contains information regarding mouse movement or click actuation. Each of such a message first goes through a filtration process to filter out any corrupted data. After filtration, mouse movement data pass through a smoothing process for noise and jitter mitigation. The current implementation of the driver performs smoothing utilizing a Gaussian moving average filter. After filtration and smoothing, both movement and click data are passed through a remapping system, which facilitates tuning of the cursor movement speed, click inversion (switching left click to right click and vice versa), and vertical movement inversion (tilting the head upward will move the mouse cursor downward and vice versa). Finally, the data are transformed into appropriate operating system calls to emulate mouse cursor movement or click actuation.

        \subsubsection{Gesture Data}
            A \textit{gesture data} message has information regarding which kind of gesture has been executed from the receiver side. Moreover, the driver supports processing of mouse movement data directly to identify pattern-based gestures as well. Once a gesture has been identified, the driver executes the corresponding action. For instance, moving the head upward while twitching both cheek muscles will disable the mouse, while moving the head downward and twitching both cheek muscles will enable the mouse. The future version of the driver is planned to include custom user gesture recognition and action mapping. 
        
        \subsubsection{Info Data}
            The \textit{info data} message, is used to convey information regarding the state of the device to the driver. Based on the info, the driver can report state changes (such as connected or disconnected states) or errors to the user.
        
        

\section{Method}
    This study was designed to evaluate and compare users’ interaction in pointing tasks while using Auxilio against a commercially available and patented camera-based head-tracking AMC, the Smyle Moouse \cite{SmyleMouse1, SmyleMouse2, SmyleMouse3}, as it only requires a generic webcam for interaction. This research was reviewed, approved, and funded by the Institutional Review Board of the affiliated university of the authors. This section begins with the details of our study participants, experimental design, followed by results analysis and discussion sections.
        
    \begin{figure}[htbp]
        \centering
        \includegraphics[width=.85\textwidth]{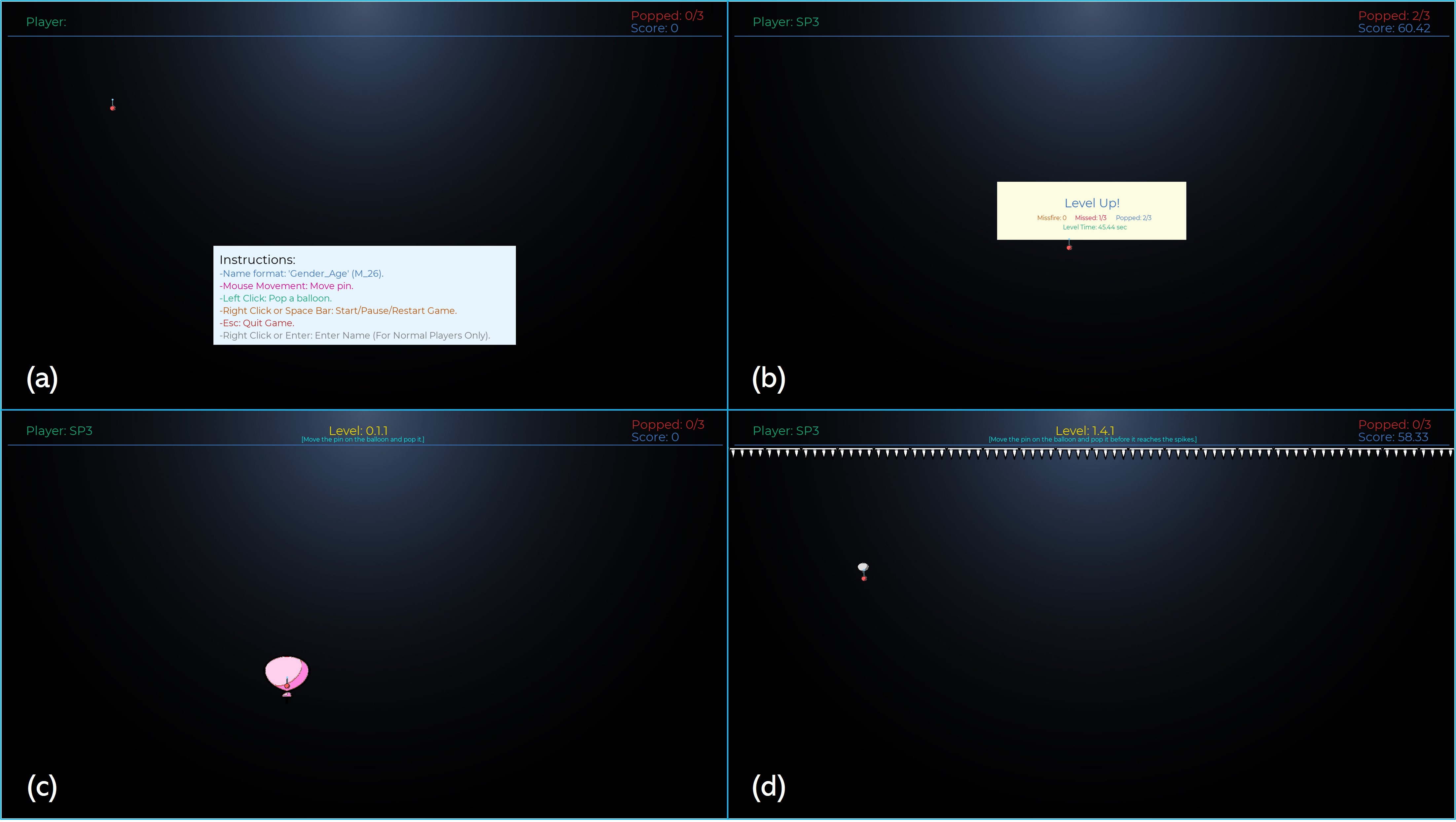}
        \caption{Snapshots of the game \textit{Popper} – (a) Player registration and game instruction screen, (b) Resting period in the form of a reward screen between levels for reducing fatigue,  a particular level with (c) a static balloon of size 128px, and (d) a dynamic balloon of size 32px.}
        \Description{A four-part figure, numbered from a to d, showing the user interface of the pointing game Popper. The interface shown in part a asks the player to enter their name, followed by instructions on the gameplay. The interface shown in part b shows a snapshot of a reward screen upon level completion, with statistics of missed clicks and time required to complete the associated level. It also contains an information pane on top showing the player's name and the number of balloons popped, ordered from left to right. The interface shown in part c depicts a particular level of the game in progress, with a corresponding snapshot of popping a floating balloon as a static target of size 128px on the screen by pointing to it using the mouse cursor. It also contains an information pane on top showing the player name, current level number and the number of balloons popped, ordered left to right. The interface shown in part d depicts a particular level of the game in progress, with a corresponding snapshot of popping a floating balloon as a moving target of size 32px on the screen by pointing to it using the mouse cursor. It also contains an information pane on top showing the player name, current level number and the number of balloons popped, ordered left to right.}
        \label{fig:Popper}
    \end{figure}
    
    \subsection{Participants} \label{subsec:participants}
\begin{figure}[htbp]
            \centering
            \includegraphics[width=0.8\linewidth]{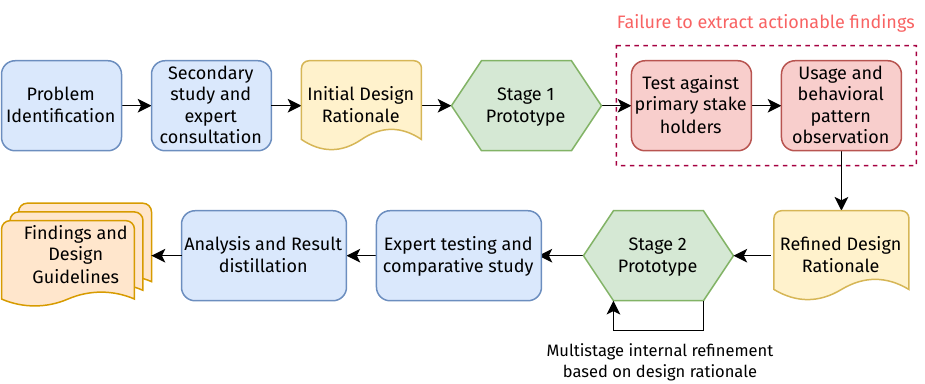}
            \caption{Overall design workflow of Auxilio.}
            \label{fig:overall_design_workflow}
            \Description{Overall workflow of designing Auxilio. It depicts the journey from a functionality-focused prototype to primary stakeholder testing. The failure then led to a refined design rationale with iterative refinement and test against experts, to reach final design guidelines.}
        \end{figure}

        Prior to our current study, we visited a local NGO and performed a study with first-hand stakeholders, i.e.\ people with upper-limb disabilities. %
        Our initial design rationale was strictly functionality oriented.
        However, based on our experience with the first-hand stake holders, we identified two major flaws with our initial study design. Firstly, considering the socio-economic context and computer literacy in our country, while it is difficult to access upper-limb disabled people who can operate a computer, it is even more difficult to find potential candidates who could provide structured constructive feedback on the device design and working mechanism. Secondly, as most of the stakeholders lived below the poverty line, it is not possible to get continuous access to them throughout the whole device design life-cycle. Hence, we had to rethink our design rationale (see \autoref{subsec:des_ration}), and felt the necessity to conduct a technical evaluation of our AMC utilizing experts with sufficient domain knowledge to provide us with structured feedback. This decision led us to finding salient technical insights related to such AMC design that the lay users could not possibly have articulated (see \autoref{sec:qualfeedback}). We recruited $10$ participants [$7$ males ($70\%$, Mean Age: $27.71 \pm 6.85$ years), $3$ females ($30\%$, Mean Age: $25.67 \pm 2.08$ years)] from known acquaintances via email, who are either educators or students in the domain of computer science and other relevant disciplines. They were notified about the automated data collection before their participation and were assured of no invasion of privacy on our part. Each participant provided written consent (see \autoref{app:consent}) before they participated in this study. As a token of appreciation for their time and efforts, a remuneration equivalent to $4.20$ USD in local currency was handed over to each of them at the end of the respective data collection session. We also collected their skin temperature, galvanic skin response and heartbeat using non-invasive wearable sensors to measure induced stress. However, the data did not reveal any actionable insight, and therefore, it was not used in the study. Our overall design approach is summarized in \autoref{fig:overall_design_workflow}.

    \subsection{Experimental Design}
        We conducted a within-subject \textit{Point and Click} experiment, featuring a balloon popping game, \textit{Popper} \cite{kabir2022antasid}, with both Auxilio and Smyle Mouse as AMCs without imposing any interaction constraints such as --- \textit{extremely accurate, accurate, neutral, fast}, and \textit{extremely fast} \cite{c111}. The experiment had only one independent variable, the device being tested --- Auxilio and Smyle Mouse. The dependent variables were the different performance metrics (see \autoref{tab:Popper data summary}).  We allotted around 35--50 minutes per participant during which they were briefed about the objective of this experiment, the semantics of the game \textit{Popper}, and the interaction mechanisms of each AMC. They had a few trial runs to familiarize themselves with the respective AMCs, followed by the experiment. 

        The experiments were performed in the same controlled environment for both the devices. Each participant played the entire game in fullscreen mode on a desktop with a screen resolution of $1920\times1080$ pixels using either Auxilio first, followed by Smyle Mouse, or vice versa. The sequence of device usage was alternated between every participant to enforce counter balancing. A brief resting period was allocated after each device interaction to reduce fatigue. The participants were presented with only one-at-a-time randomly-appearing $36$ static and $36$ moving or dynamic balloons of $4$ different widths ($32$ px, $64$ px, $96$ px, and $128$ px) as targets that they had to pop using a left mouse click inside it with respective AMCs. For the dynamic targets, a balloon would float upwards on the screen until it reached several spikes placed on top, causing the balloon to pop if the player failed to do so. We term each pop of a balloon as a \textit{trial}. A \textit{trial} began with the target appearance and ended with a successful pop action. In summary, Each participant was presented with $72$ targets of various widths in total, half of which were static and the rest were dynamic. The game ended when all the targets were popped by a participant. The corresponding interaction data were collected for both devices per participant. A few snapshots of \textit{Popper} have been provided in \autoref{fig:Popper}. 
        
        Following their interaction with the devices, each participant was asked to share their perceptions of the usability, the pros, and the cons of both AMCs through an unstructured interview, audio recordings of which were collected and analyzed qualitatively later. Finally, they were asked to complete the SUS questionnaire \cite{c60, c61, c62, c63, c64, c65} (\autoref{app:SUS items}) online to rate the various usability aspects of each AMC. Their responses (\autoref{app:SUS ratings_auxilio} and \autoref{app:SUS ratings_smyle}) were later processed to analyze the usability of Auxilio in comparison with Smyle Mouse. It is important to note that we carried out the usability survey after the interview session to prevent the participants' open opinions regarding their interaction with the AMCs from being influenced by the SUS questionnaire. This allowed us to get unbiased and natural insights into their experience with the AMCs.


    \subsection{Data Collection and Analysis}
        The data collected for analysis in this study consists of the participants' interaction data from the game \textit{Popper} for each AMC (Auxilio and Smyle Mouse) as a pointing device, the audio recordings of their unstructured interviews, and their responses to the SUS questionnaire. The interaction data per participant consists of parameters such as --- \textit{movement amplitude} ($A$), \textit{target width} ($W$), \textit{movement time} ($MT$), \textit{cursor on and off count per target}, \textit{straight-line distance ($SLD$) between the cursor and the target coordinates}, \textit{total mouse clicks}, \textit{total miss-clicks}, \textit{total targets hit}, and \textit{total targets missed}. Given a particular type of target, the parameter \textit{cursor on and off count per target} was collected to get a measure of Target Re-entry (TRE) for each AMC, which is defined as the event when a mouse cursor enters the target region, then leaves, and re-enters again \cite{mackenzie2001accuracy}. The interpretations of the recorded parameters are summarized in \autoref{tab:Popper data summary}. 
                    
        \begin{table}[htbp]
        \centering
        \small
        \caption{Descriptive summary of the parameters collected per participant in the  \textit{Point and Click} experiment using Auxilio \cite{kabir2023acceptability} and Smyle Mouse \cite{SmyleMouse1, SmyleMouse2, SmyleMouse3}.}
        \label{tab:Popper data summary}
            \begin{tabularx}{\textwidth}{lcl}
                \toprule
                 \parbox{40ex}{\centering \textbf{Parameter}} & \parbox{10ex}{\textbf{Value(s)/Unit}} & \parbox{68ex}{\centering \textbf{Interpretation}}\\ 
                \midrule
                \textit{Movement Amplitude} ($A$)& pixels &\parbox{68ex}{Represents the length of the cursor trajectory as the movement amplitude ($A$) in \autoref{eq:ID}.}\\[1em]
                \textit{Target Width} ($W$)& pixels &\parbox{68ex}{Represents the width of a target ($W$) in \autoref{eq:ID}.}\\
                \textit{Movement Time} ($MT$) & seconds &\parbox{68ex}{Represents the time ($MT$) required to hit a target in \autoref{eq:TP}.}\\[0.5em]
                \textit{Cursor On and Off Count Per Target}& $[I, O]$ &\parbox{68ex}{The total number of times the cursor moved inside ($I$) and outside ($O$) a target's boundary.}\\[1em]
                \parbox{40ex}{\textit{Straight-line Distance ($SLD$) between the Cursor and the Target Coordinates}} & pixels &\parbox{68ex}{The Euclidean distance between the coordinates of the mouse cursor and the target.}\\[1em]
                \textit{Total Mouse Clicks*} & $n$ &\parbox{68ex}{The total number of mouse clicks.}\\
                \textit{Total Miss Clicks*} & $n$ &\parbox{68ex}{The total number of mouse clicks outside a target boundary.}\\
                \textit{Total Targets Hit*} & $n$ &\parbox{68ex}{The total number of targets (balloons) that were \textit{popped}.}\\
                \textit{Total Targets Missed*} & $n$ &\parbox{68ex}{The total number of targets (balloons) that were \textit{missed}.}\\
                \bottomrule
                \multicolumn{3}{l}{* Total value at the end of the entire game \textit{Popper}.}
            \end{tabularx}
            
        \end{table}

        For a comparative evaluation of a user's performance with different AMCs in pointing tasks with different indexes of difficulties ($ID$), the index of performance, also known as throughput ($TP$) \cite{c114, kabir2022antasid} is usually considered \cite{cicek2020designing}. Once the data from this experiment for both AMCs were collected, we calculated the $ID$s for each task using \autoref{eq:ID}, where $A$ is the movement amplitude and $W$ is the target width \cite{kabir2022antasid}. The $TP$ of each participant in pointing tasks with static and dynamic targets, measured in \textit{bits per second} (bps), were separately calculated as the average ratio of $ID$ and $MT$ over $n$ pointing tasks \cite{c114, kabir2022antasid} (\autoref{eq:TP}). We also calculated the path efficiency ($PE$) of a the cursor trajectory as the ratio of $SLD$ and the movement amplitude ($A$) (\autoref{eq:PE}). Then we carried out a one-tailed unpaired $t$-test ($\alpha=0.05, 9$) between the users' throughput with each AMC to analyze if the mean throughput of one is significantly higher than the other given a particular type of target (static or dynamic). We also calculated and analyzed the \textit{miss click rates}, \textit{target hit/miss rates for dynamic targets}, \textit{mean target re-entry}, and \textit{mean path efficiency} for each pointing device (Auxilio and Smyle Mouse) to get an insight into the individual and/or the combined effect of the precision and stability of the cursor movement, and the effectiveness of the click mechanisms of each AMC on user performance for pointing at a target (static or dynamic) of a particular width (32px, 64px, 96px, and 128px). 
                
        \begin{align}
            ID &= \log_{2}\left(\frac{A}{W} + 1 \right)\label{eq:ID}\\
            TP &= \frac{1}{n}\sum_{i=1}^{n}\frac{ID_i}{MT_i} \label{eq:TP}\\
            PE &= \frac{SLD}{A} \label{eq:PE}
        \end{align}

        Next, we transcribed the audio recordings of the unstructured interview to conduct a qualitative analysis to capture the strengths and weaknesses of both the devices from the participants' point of view. A hybrid coding approach (deductive and inductive) was employed, with two raters discussing and reconciling differences in coding. We set some predefined codes based on the interviews and our prior experience. After dividing the transcripts into taggable units, two raters independently coded all the units. While coding, the raters were allowed to create any emerging code if necessary. After all the transcripts had been tagged, reconciliation of emerging codes was performed between the two raters, followed by the calculation of Cohen's Kappa \cite{cohen1960coefficient} measure for inter-rater reliability. Finally, based on the agreement of both the raters, final tagged versions of the transcripts were produced. After that, the codes were merged to find out the major themes for each device based on the frequency of codes. Future suggestions for each device were also extracted from the transcripts. We concluded our experiment with the analysis of participants' responses to the SUS questionnaire \cite{c60}.
        
    \section{Results and Discussions}
        \subsection{Point and Click Experiment}
            The descriptive statistics of user interaction data offers insights into usability and user performance with Auxilio and Smyle Mouse as AMCs (see \autoref{tab:Popper desc stat}). A two-tailed $F$-test revealed that the throughput varies insignificantly ($\alpha=0.05$) for both static and dynamic targets for both devices ($F_{Static}(9,9)=1.9755, p=0.3250$; $F_{Dynamic}(9,9)=3.6096, p=0.0694$). However, a one-tailed unpaired $t$-test ($\alpha=0.05$) revealed that the mean throughput for dynamic targets with Auxilio ($\overline{TP}_{Auxilio}=0.9315\;bps$) was significantly higher than that with Smyle Mouse ($\overline{TP}_{Smyle\;Mouse}=0.6654\;bps$) ($t_{Dynamic}(9)=3.3641, p=0.0024$), while it was insignificant for static targets\;($\overline{TP}_{Auxilio}=0.8671\;bps$, $\overline{TP}_{Smyle\;Mouse}=0.7489\;bps$; $t_{Static}(9)=1.5340, p=0.0723$).

        \begin{table}[htbp]
            \centering
            \caption{Descriptive statistics of the task completion times and throughput in the \textit{Point and Click} experiment using different pointing devices.}
            \label{tab:Popper desc stat}
            
            \begin{tabular}{cccccccccc}
                \toprule
                \multirow{2}{*}{\textbf{\parbox{9ex}{\centering Target\\Category}}} & \multirow{2}{*}{\textbf{\parbox{8ex}{\centering Pointing\\Device}}} & \multirow{2}{*}{\textbf{\parbox{5ex}{\centering Tasks ($n$)}}} & \multicolumn{4}{c}{\textbf{\parbox{34ex}{\centering Task Completion Time (seconds)}}} && \multicolumn{2}{c}{\textbf{\parbox{15ex}{\centering \textbf{$\textbf{TP}$ (bps)}}}}\\ 
                \cline{4-7} 
                \cline{9-10}
                 &  & &\textit{\parbox{5ex}{\centering Mean}} & \textit{\parbox{3ex}{\centering SD}} & \textit{\parbox{3ex}{\centering Min}} & \textit{\parbox{3ex}{\centering Max}}  && \textit{\parbox{5ex}{\centering Mean}}  & \textit{\parbox{10ex}{\centering t(0.05, 9)}}\\ 
                 \midrule
                \multirow{2}{*}{Dynamic} & Auxilio & 360 & 8.5272 & 9.9313 & 0.7970 & 59.063 && 0.9315 &\multirow{2}{*}{\parbox{10ex}{\centering \textbf{0.0024}*}}\\
                                         & Smyle Mouse & 360 & 13.0633 & 12.2990 & 1.1720 & 60.9690 && 0.6654 & \\[0.5em]
                \multirow{2}{*}{Static} & Auxilio & 360 & 6.4358 & 5.8687 & 0.5780 & 46.5000 && 0.8671 & \multirow{2}{*}{\parbox{10ex}{\centering 0.0723}}\\
                                        & Smyle Mouse & 360 & 8.3471 & 9.6888 & 1.2820 & 115.063 && 0.7489 & \\

                \bottomrule
                \multicolumn{10}{l}{* $p$-value is significant at $\alpha=0.05$.}
            \end{tabular}
        \end{table}
        
        The participants reported difficulty in simultaneously pointing and clicking small targets (both static and dynamic) with Auxilio (due to cursor oscillations even without head movements) compared to Smyle Mouse (due to discrete, stable cursor movements). This oscillation increased their \textit{miss-click rates}, \textit{mean target re-entry}, and in some cases their task completion times. However, no significant inconsistencies in click detection were reported for Auxilio. On the contrary, Smyle Mouse failed to recognize mouse click events frequently, inflating their task completion time (both static and dynamic targets) and target miss rates (dynamic targets only).
        
        \begin{figure}[htbp]
            \centering
            \begin{subfigure}{0.43\textwidth}
                \includegraphics[width=\textwidth]{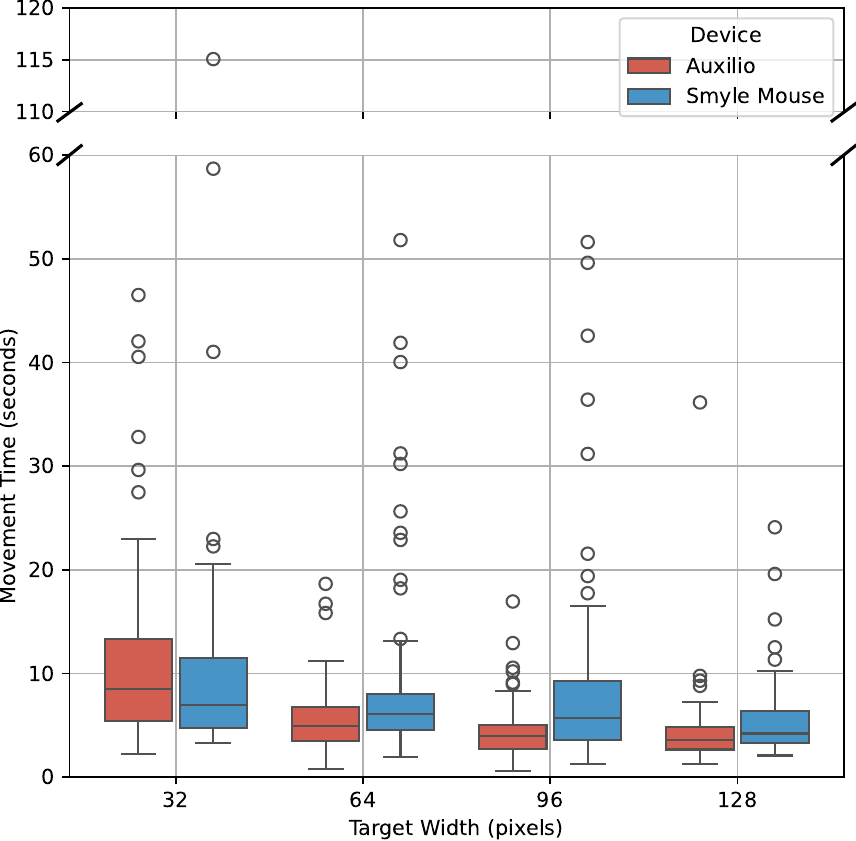}
                \caption{Time required to hit static targets}
                \label{fig:boxplot static pops}
            \end{subfigure}
            \hspace{3em}
            \begin{subfigure}{0.42\textwidth}
                \includegraphics[width=\textwidth]{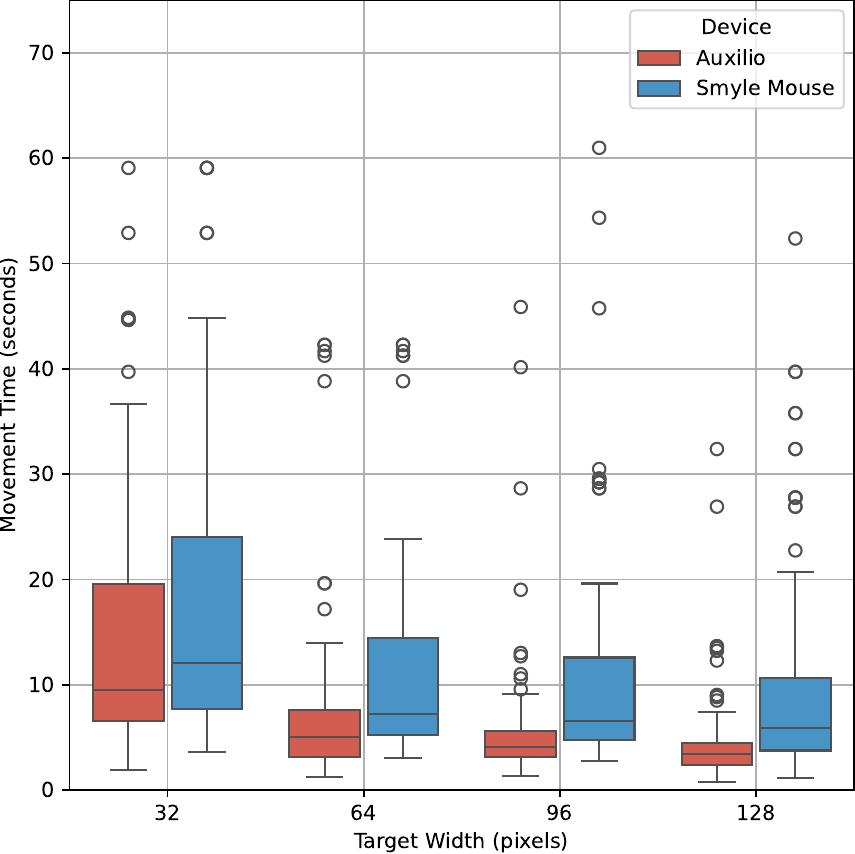}
                \caption{Time required to hit dynamic targets}
                \label{fig:boxplot dynamic pops}
            \end{subfigure}
                    
            \caption{Box plots of the time required to hit static and dynamic targets with Auxilio and Smyle Mouse grouped by target width.}
            \Description{A two-part figure showing the boxplots of task completion time using Auxilio and Smyle Mouse for static targets in part a and dynamic targets in part b grouped by target width. }
            \label{fig:poptime analysis popper}
        \end{figure}
        
        \begin{figure}[htbp]
            \centering
            \begin{subfigure}{0.48\textwidth}
                \includegraphics[width=\textwidth]{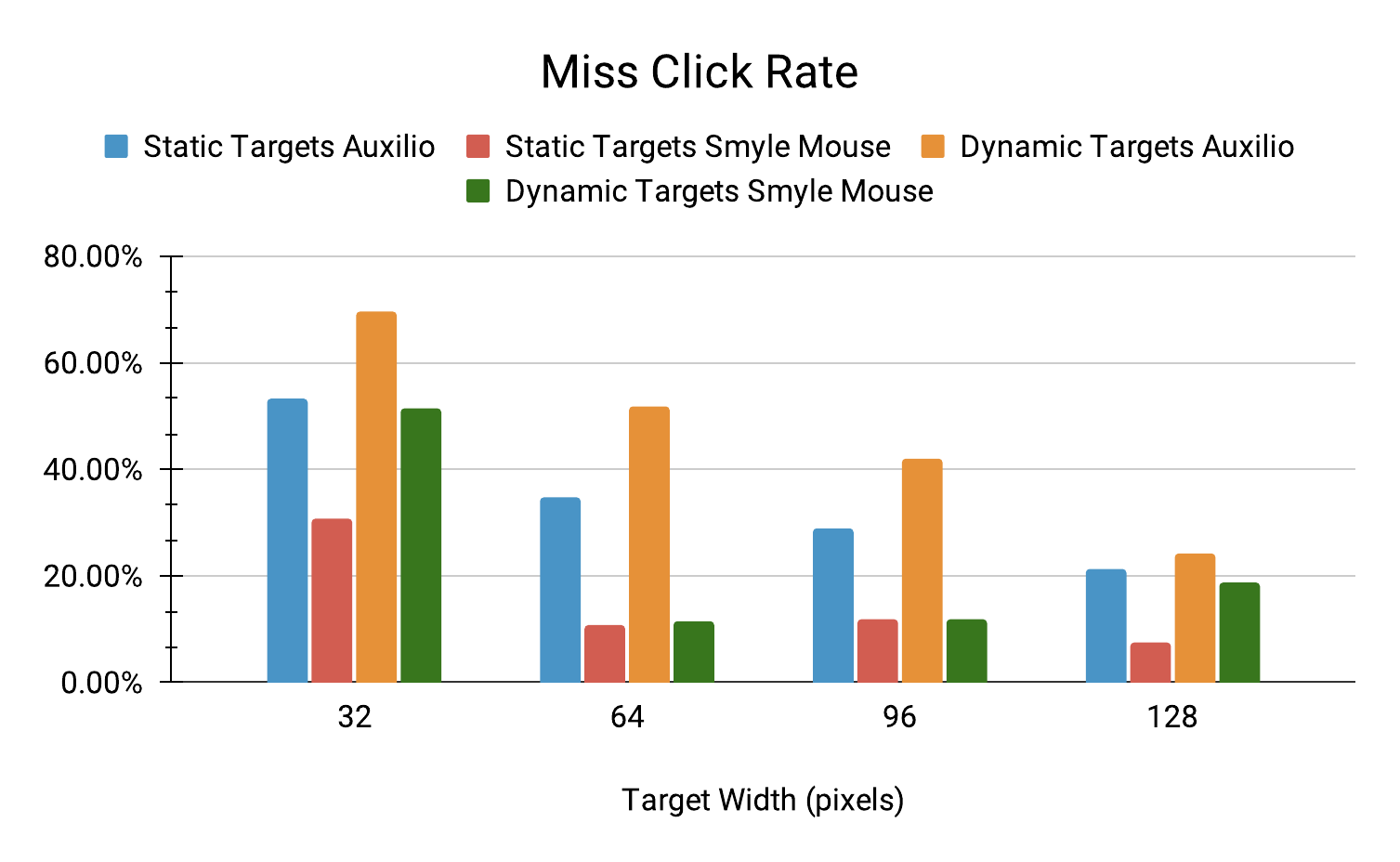}
                \caption{}
                \label{fig:miss click}
            \end{subfigure}
            \hfill
            \begin{subfigure}{0.48\textwidth}
                \includegraphics[width=\textwidth]{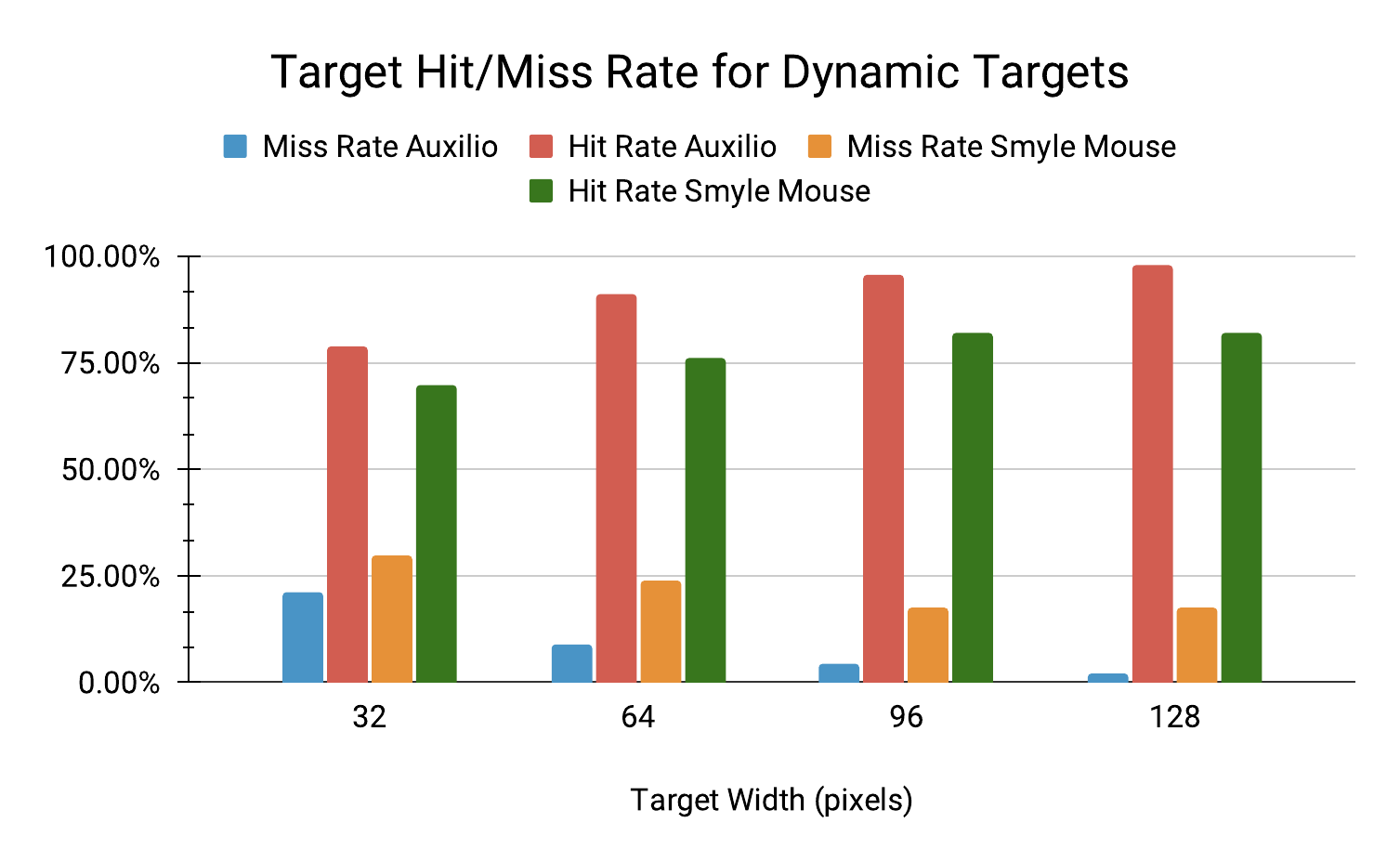}
                \caption{}
                \label{fig:target hit_miss rate}
            \end{subfigure}
            \hfill
            \begin{subfigure}{0.48\textwidth}
                \includegraphics[width=\textwidth]{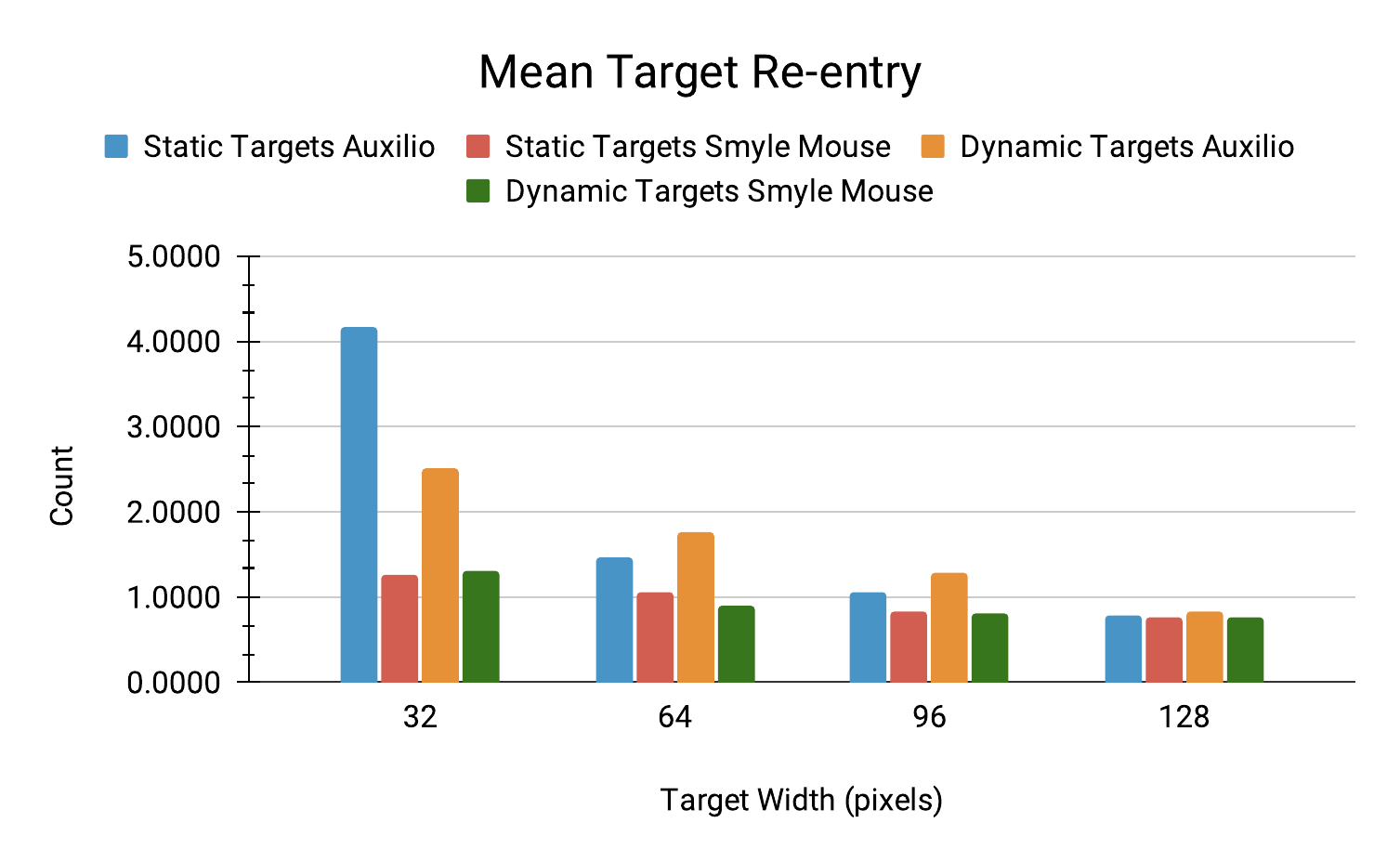}
                \caption{}
                \label{fig:mean tre}
            \end{subfigure}
            \hfill
            \begin{subfigure}{0.48\textwidth}
                \includegraphics[width=\textwidth]{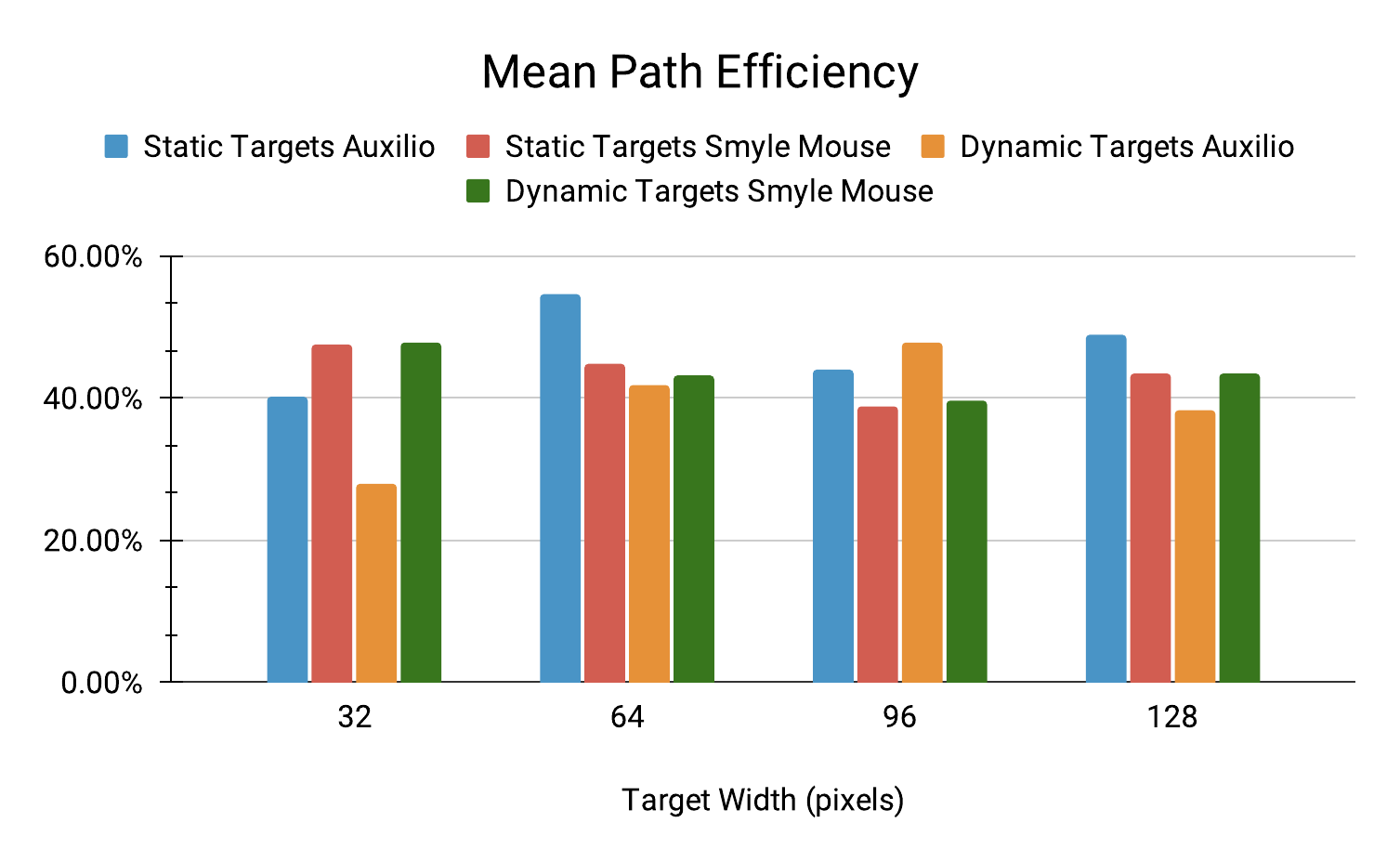}
                \caption{}
                \label{fig:mean pe}
            \end{subfigure}
                    
            \caption{User performance analysis in terms of --- (a) Miss Click Rate (lower value indicates stable cursor control and high value indicates sensitive click actuation), (b) Target Hit/Miss Rate (higher hit rate indicates better control), (c) Mean Target Re-entry (lower value indicates stable cursor control), and (d) Mean Path Efficiency (higher value indicates better control), with each pointing device (Auxilio and Smyle Mouse) grouped by target width for different types of targets (Static and Dynamic).}

            \Description{A four-part figure numbered from a to d showing the results of the parameters used for user performance analysis with each pointing device (Auxilio and Smyle Mouse) grouped by target width for different types of targets (Static and Dynamic). part a shows the Miss Click Rate where a lower value indicates stable cursor control and high value indicates sensitive click actuation. part b shows Target Hit/Miss Rate where a higher hit rate indicates better control. part c shows Mean Target Re-entry where a lower value indicates stable cursor control. finally part d shows Mean Path Efficiency where a higher value indicates better control.}
            \label{fig:performance analysis popper}
        \end{figure}

        The wider range of \textit{task completion times} (\autoref{fig:poptime analysis popper}) and higher values of \textit{miss click rates} (\autoref{fig:miss click}), \textit{dynamic-target hit/miss rates} (\autoref{fig:target hit_miss rate}), and \textit{mean target re‑entry} (\autoref{fig:mean tre}) for 32-pixel static targets with Auxilio substantiate the reported cursor oscillation, hindering precise clicking. A decreasing trend in these parameters with increasing target width for both target types suggests that the cursor oscillation can be compensated with targets of optimal size. On the contrary, a reduced \textit{mean target re-entry} ($< 1.5$) for Smyle Mouse (\autoref{fig:mean tre}) confirms its stable cursor movement. However, its task completion times for static targets were inconsistently high, exceeding $10$ seconds for $128$-pixel targets and even reaching $115$ seconds for $32$-pixel targets (\autoref{fig:boxplot static pops}), validating the reported frequent failures in click-detection. Despite cursor oscillations, Auxilio featured a reliable click-detection mechanism, as reflected in the decreasing trend of \textit{miss-click rate} with increasing target width for both static and dynamic targets, further evident from subsequent qualitative analysis. Interestingly, the \textit{mean path efficiency} of both AMCs never exceeded $55\%$ regardless of type and width of targets. The involuntary oscillations in Auxilio and the discrete cursor movements in Smyle Mouse may have contributed to this effect, as seen from the movement heatmaps ( \autoref{app:heatmaps}). To summarize, Auxilio demonstrated substantial potential to outperform Smyle Mouse in regular use cases, except for very small targets.

    \subsection{Feedback and Observation} \label{sec:qualfeedback}
    The qualitative analysis of the participant interviews helped us find qualitative insights into each of the devices. After reconciling the inductive codes, Cohen's Kappa measure was found to be 0.72 for around thirty codes in each category, indicating a substantial inter-rater reliability \cite{cohen1960coefficient}. Summary of the themes, codes, and their respective frequencies for Auxilio and Smyle Mouse are presented in \autoref{app:themescodes_auxilio} and \autoref{app:themescodes_smyle}, respectively. 
    
    The analysis aimed to complement the experimental findings by identifying key themes related to user experiences, strengths, and weaknesses of each of the devices, followed by potential design directives for similar devices in the future. In the case of Auxilio, the analysis revealed six prominent themes ---
    
    \begin{enumerate}
        \item \textbf{Ease of Use and Deterministic Behavior:} One of the main strengths of Auxilio comes from its ease of use and persistent and easy calibration. One of the participants, P4 reported, ``The calibration in Auxilio was easy, user friendly, and consistent throughout the session.'' Another major strength is the deterministic behavior of Auxilio, enabling participants to infer the state of the system at any moment and act accordingly, as reported by P2, ``Both devices required neck movements. For Auxilio, neck movements are certain, I know exactly how much to move. But I could not map myself in the case of Smyle Mouse.''
        \item \textbf{Difficult Precise and Static Pointing:} The main weakness of Auxilio comes from perhaps the inability to keep the cursor static at a single point as it is directly bound to the head movements of the user, which results in small oscillations and difficulty while trying to hit small targets. According to P2, ``The mouse cursor had jittery movements while trying to pinpoint on the smaller (32px) targets. It was in the periphery, but it was still moving. Pinpointing and clicking on even smaller targets would have been a problem.''
        \item \textbf{Ergonomics, Cognitive Load, and Comfort:} The overall design of Auxilio was mostly appreciated by the users with a few suggestions discussed later. As stated by P7, ``I did not face any strain while moving my head using Auxilio'', and P2, ``Mental load was lower in Auxilio, as the pointer movement was bound to exact head movements.''
        \item \textbf{High Responsiveness:} As indicated by the experimental results, the main reason Auxilio performed better than Smyle Mouse in the dynamic target tasks is due to its high responsiveness for both clicking and mouse cursor movements. In this regard, participant P8 stated, ``In Auxilio, after the mouse calibration, the cursor always returned to the screen center when I wanted to recenter it. The cursor followed whichever direction I intended to move it with my head movement. I never had to move my head to an extreme angle to move the cursor with Auxilio.''
        \item \textbf{Adaptability:} Based on the interviews of some of the participants, Auxilio, specially for clicking, had some initial learning curves. However, it was easy to adapt as all the participants got accustomed to it by the end of their respective sessions. Based on the statement of P2, ``Clicking in Auxilio at the beginning required coping up with cheek muscle movement. I had to practice it a couple of times. After learning, it was ok.''
        \item \textbf{Discomforts:} Few participants reported some discomfort while using Auxilio. The majority of the comments were about the large visor of the device, which in some cases, obstructed part of their peripheral vision, creating uneasiness. Participant P3 reported, ``However, the current design of the device features a visor-like mechanism which I felt blocked my view sometimes.''
    \end{enumerate}
    
    On the other hand, analysis of the interviews for Smyle Mouse pointed to four prominent themes ---
    
    \begin{enumerate}
        \item \textbf{Performance and Calibration Issues:} The greatest issue of Smyle Mouse was with its inconsistent performance and the requirement of frequent recalibrations. Despite the initial calibration of the mouse cursor, it lost mapping with the face orientation over time and the participants were continuously struggling to make up for that. As a result, patterns of continuously dragging the cursor towards the different screen edges can be seen from the heatmaps in \autoref{app:heatmaps}. According to P8, ``Using the Smyle Mouse was a bit stressful for me because, if I moved my head to extreme left/right, the cursor would not return to the exact center. To recenter the cursor, I had to move my head very fast in the opposite direction.'' The smile detection algorithm of the Smyle Mouse is also prone to drifts and it is inaccurate. It caused the greatest dissatisfaction among the users and combined with the cursor drift and dynamic targets, it resulted in serious difficulties in detecting clicks after a moment and became almost impossible to invoke clicks from extreme angles. According to P6, ``I felt the calibration of the Smyle Mouse drifted overtime during the session; it did not register click events.'' Another complaint was about dynamic target tracking. Smyle Mouse relies on discrete movements to keep the cursor static at a point, as seen from the movement heatmaps in \autoref{app:heatmaps}. However, the same design choice made it difficult to continuously track dynamic targets. According to P1, ``But it can also become a problem when you are trying to do small movements using your mouse, specially for dynamic target tracking, when you need to make small adjustments to your cursor position. This is difficult to do because of how the cursor moves (discrete movements).'' As a whole, it resulted in a non-deterministic and inconsistent system.
        \item \textbf{Ergonomic Issues, Cognitive Load, and Physical Discomforts:} The overall performance problems with Smyle Mouse translated to ergonomic issues and higher cognitive load. It required a substantial amount of head and neck (for cursor) and jaw muscle movement (for clicking via smile), resulting in increased physical stress. P8 commented, ``Due to the rapid movement of my head, although I did not experience any pain, my neck felt stiff after using the Smyle Mouse for a while which was stressful. Due to the fact that I had to move my head to an extreme angle frequently while using the Smyle Mouse, it was difficult to look at the screen from that angle.'' Moreover, the continuous struggle to re-calibrate the cursor and make up for inconsistency resulted in a higher cognitive load as well. In this regard, P2 stated, ``The problem with the Smyle Mouse was that it required calibration before every click, resulting in higher mental load.''
        \item \textbf{Design and Learnability Issues:} Relying on a camera-based setup makes the Smyle Mouse inherently difficult to control at extreme angles and works properly only within a constrained area. However, the issue with drifting makes the mouse continuously succumb to this problem. According to P9, ``I had to move my head to an extreme angle when I wanted to move the cursor to an extreme point on the screen occasionally. The Smyle Mouse failed to recognize my smile.'' Along with that, the initial calibration of the Smyle Mouse was reported as difficult to understand, resulting in learnability issues. However, few participants did find the clicking and calibration of the Smyle Mouse to be easy. 
        \item \textbf{Stable Static Pointing:} While the Smyle Mouse suffers from many issues, it has one advantage over Auxilio. Due to the discrete nature of mouse pointer movement, it was easier to hit smaller (32px) static targets with the Smyle Mouse compared to Auxilio as depicted by the lower value of mean task completion time in \autoref{fig:poptime analysis popper}. In this regard, P7 commented, ``In the case of the Smyle Mouse, I felt that pinpointing was more accurate as the cursor did not have any jittery movement.''
    \end{enumerate}

    \subsection{System Usability}\label{sec:sus}
        In this analysis, we used the System Usability Scale (SUS) \cite{c60} to understand users' perspective on the usability of Auxilio in comparison with Smyle Mouse. SUS employs a $10$-item questionnaire to quantify the subjective assessments through a usability score between $0$ and $100$. To meaningfully interpret these scores (\textit{poor} or \textit{good}, letter grades (A+, A, etc.) are assigned to them, $A+$ being the highest grade of usability \cite{c60, c133, c134}.
        
        As seen from \autoref{tab:SUS ratings}, the SUS scores for Auxilio ranged from $70.00$ to $92.50$ (Mean: $78.75$), corresponding to a grade of $B+$ and an adjective usability rating of \textit{Good} \cite{bangor2009determining}. On the other hand, the scores for Smyle Mouse ranged from $30$ to $92.50$ (Mean: $48.75$), resulting in a grade of $F$ and an adjective rating of \textit{Awful}.

        The perceptions on the usability of the devices varied among participants. Only $10\%$ expressed their interest in using Smyle Mouse frequently, while $10\%$ expressed reluctance to use Auxilio frequently. All participants agreed on the simplicity of the interaction mechanism of Auxilio, whereas $70\%$ of them disagreed for Smyle Mouse. Details of the participants' responses to the SUS questionnaire have been provided in \autoref{app:SUS ratings_auxilio} and \autoref{app:SUS ratings_smyle}.
        
    \begin{table}[htbp]
            \centering
            \caption{Results of usability analysis of Auxilio and Smyle Mouse \cite{SmyleMouse1, SmyleMouse2, SmyleMouse3} using the System Usability Scale (SUS) \cite{c60}.}
            \label{tab:SUS ratings}
            \begin{tabularx}{.59\textwidth}{cccccc}
                \toprule
                \multirow{2}{*}{\parbox{15ex}{\centering \textbf{Respondent}}} & 
                \multicolumn{2}{c}{\centering \textbf{Auxilio}} && 
                \multicolumn{2}{c}{\centering \textbf{Smyle Mouse}}\\
                \cline{2-3}
                \cline{5-6}
                & \textbf{SUS Score} & \textbf{Grade} && \textbf{SUS Score} & \textbf{Grade} \\ 
                \midrule
                \textbf{R1} & 92.50 & A+ && 92.50 & A+\\
                \textbf{R2} & 85.00 & A+ && 82.50 & A\\
                \textbf{R3} & 85.00 & A+ && 40.00 & F\\
                \textbf{R4} & 87.50 & A+ && 32.50 & F\\
                \textbf{R5} & 52.50 & D && 45.00 & F\\
                \textbf{R6} & 70.00 & C && 32.50 & F\\
                \textbf{R7} & 82.50 & A && 65.00 & C\\
                \textbf{R8} & 70.00 & C && 30.00 & F\\
                \textbf{R9} & 72.50 & C+ && 30.00 & F\\
                \textbf{R10} & 90.00 & A+ && 37.50 & F\\
                \midrule
                \textbf{Mean} & 78.75 & \multirow{2}{*}{B+} && 48.75 & \multirow{2}{*}{F} \\
                \textbf{SD} & 12.32 & && 23.01 & \\ \bottomrule
            \end{tabularx}%
    \end{table}


    \subsection{Design Guidelines}
    Based on our experience with AMC designing and the pointing experiment, system usability, and qualitative analysis of the strengths and weaknesses of both devices in \autoref{sec:qualfeedback}, we outline several guidelines specific to the design of similar AMCs in the future, which are grounded on experimental and empirical data, and not evident from common design principles. These guidelines will also serve as the basis of our updated design rationales.

    \begin{enumerate}
        \item \textbf{Tunable Mapping --- Balancing Absolute vs Relative Positioning:} As seen from the studies, Auxilio demonstrates superior performance in dynamic target acquisition due to its absolute cursor mapping. However, the same makes it difficult to keep the pointer stable on small static targets, introducing jitter. On the other hand, while Smyle Mouse performs poorly in dynamic tracking due to its relative mapping scheme, the same allows it to stabilize the cursor on static targets. Therefore, either a designer needs to choose the mapping strategy based on the type of intended tasks to be performed by the AMC or for robust general use cases, provide adjustable mapping algorithms. A user should be able to toggle between ``Absolute Mode’’ for high-speed tasks (e.g., gaming, scrolling) and ``Relative Mode’’ for precision tasks (e.g., text selection) on the fly (suggested: based on gestures) as no single mode suits both. It should be added that smoothing algorithms (e.g., different low-pass and moving average filters) can stabilize the cursor even in absolute mapping, but that comes at the cost of reduced responsiveness.
        \item \textbf{Explicit Clutching for Drift Recovery:} Auxilio, designed to work purely in absolute mode, has a deterministic center for the cursor and allows the user to control its movement while always being within a fixed zone. On the contrary, Smyle Mouse, being dependent on vision based relative head (via nose) tracking, naturally suffers from the drift of the user's comfortable ``center’’ over time, gradually making the system less deterministic. If a designer wishes to leverage relative mapping in an AMC, it must include a rapid, gesture-based re-centering or clutch mechanism (similar to lifting a physical mouse). A gesture-based toggling (nodding or holding head at an angle) like Auxilio’s can be extended to easily accommodate this requirement, preventing the ``Gorilla Arm’’ like fatigue of holding one’s head at extreme angles. Moreover, such a mechanism also frees the users from being restricted to keeping their heads fixed at the screen with rapid connection/disconnection, allowing them to interleave everyday interaction with mouse operation freely.
        \item \textbf{Decoupled Interaction Channels to Minimize Heisenberg Effects:} A major issue in AMCs is that the act of clicking often moves the cursor or vice versa, pointing towards the Heisenberg effect. Smyle Mouse requires smiling to click, which moves the cheek and jaw, inadvertently shifting the head, thereby the cursor. Auxilio separates the channels for these effectively: the neck muscles controls the cursor and independently, cheek muscles control the clicks, minimizing cross-talk. To prevent such misclicks, the muscle group used for clicking must be mechanically isolated from the muscle group used for pointing while designing the mouse event actuations. Moreover, relying on specific muscle movement detection using complex pattern analysis, such as jaw movement for clicks (like smiling) can inherently degrade pointing precision over time due to the natural variability in the voluntary movement of such muscles.
        \item \textbf{Immediate Sensory Feedback for Non-Tactile Clicks:} Unlike a physical mouse, muscle movement based click actuation offers no tactile feedback. Smyle Mouse users have often struggled to recognize if and when a certain click action was recognized. Hence, either the click detection system needs to be extremely responsive to properly recognize every single click event, or provide visual (e.g., cursor pulse), auditory (e.g., soft click sound), or haptic feedback (body mounted vibrator) as confirmation of the down- and up-click events. Allowing the user to choose among the feedback options can broaden the operating environment (e.g., public place, areas with loud noise) of the AMC.
    \end{enumerate}

    Along with the guidelines derived from the device analysis, we also received some valuable feedback from the users about device improvements. Focusing on users suffering from neck pain, P2 suggested, ``Another possible area of improvement can be for people suffering from neck pain or disabilities, such as cervical disk prolapse. Their movement can be stiff in a specific direction. So, personalization or adaptation in the very beginning with the help of a training can be used to map the cursor. It can be adjusted and learned to improve based on performance. Having an adjustment factor could be helpful. After the initial adjustment, there could be further adjustments on-the-fly, while using the device.'' This suggestion further strengthens our very first guideline. Another common suggestion was to test the devices in prolonged use cases. During the interview, P1 emphasized, ``Having a longer session spanning hours might be more accurate to test out whether it would be viable for prolonged use or not.''
    
    One more suggestion for such devices would be to introduce different gestures as shortcuts for commonly used actions, such as connecting or disconnecting the device. Furthermore, the drivers for the software should have customization options to tune the cursor speed, swapping left and right clicks, etc. While Auxilio has these abilities, the Smyle Mouse could not support such features properly because of its design.

\section{Limitations and Future Works}

    The study was limited to basic pointing and clicking tasks, which, while important and the prime part of pointer operation, represent a subset of the full application potential of a modern computer mouse. Tasks like dragging, dropping, scrolling, and zooming were not evaluated but are key to understanding the overall capability of these devices. Future research should explore these additional tasks and how they impact user experience. Additionally, an AMC might require specialized gestures as shortcuts for common activities to improve the overall user experience. Design and evaluation of such gestures can be a significant topic of study.
    Furthermore, we did not assess the long-term and extensive use of the devices. Understanding how these AMCs perform over extended periods, including their impact on learnability, memorability, and user comfort and fatigue, is crucial. Longitudinal studies would help reveal how well the devices hold up under prolonged use and whether any issues arise over time.
    Additionally, our study focused only on AMCs developed for users with upper-limb disabilities with retained motor control over head-neck and facial muscles. However, many users may have partial head control or multiple or coexisting disabilities that could affect how an AMC is used. Future studies should consider users with other upper-limb disabilities, such as ALS or cerebral palsy, to ensure the design of AMCs is more inclusive and accessible to a wider range of users.
    
    Finally, as assistive technologies evolve, there is an opportunity to explore how AMCs could use artificial intelligence or machine learning to adapt to individual user preferences. Future work could investigate how these devices might personalize their functions over time, improving both user satisfaction and performance.
    Addressing these limitations will help develop more comprehensive design guidelines for Auxilio and similar head-mounted AMCs, making them more effective and accessible for a diverse range of users in everyday contexts.

\section{Conclusion}
    In this study, after delineating the design rationale and implementation of Auxilio, a sensor-based head-mounted mouse, we compared its performance against Smyle Mouse, a commercially available, patented vision-based device. Both devices offer similar input modalities and are designed for individuals with upper limb disabilities. Our pointing experiment showed that Auxilio generally outperformed Smyle Mouse, except when targeting the smallest static objects, proving its feasibility and highlighting the areas for improvement in Auxilio’s design. Auxilio also received a higher score on the System Usability Scale (SUS), suggesting greater practical usability. Finally, both of these analyses combined with the qualitative study of the participant interview transcripts allowed us to find out the strengths and weaknesses of the devices from the perspective of the users. By further distilling this knowledge, we were able to propose potential design guidelines and future directives for creating better AMCs for the upper limb disabled community. We believe that our effort would inspire and assist further research and innovation in developing more effective and inclusive assistive devices for this community.

\begin{acks}
    This research was reviewed and approved by \textit{Research, Extension, Advisory Services and Publications (REASP), Islamic University of Technology (IUT)} and funded by \textit{Islamic University of Technology Research Seed Grant} (IUT-RSG) under the Grant: \textbf{REASP/IUT-RSG/2021/OL/07/012}. The authors would like to thank the concerned participants for cooperating with us in conducting this research. The authors declare no conflict of interest that might have altered the course of this study in any manner.
\end{acks}

\bibliographystyle{ACM-Reference-Format}
\bibliography{Auxilio_Citations}

\pagebreak
\appendix
\raggedbottom

\section{SUS Questionnaire}\label{app:SUS items}
    \begin{table}[htbp]
        \centering
        \begin{tabular}{cp{0.8\textwidth}}
            \toprule
            \small
            \textbf{Item} & \textbf{Description}\\
            \hline
            \textbf{SUS 1} & I think that I would like to use this system frequently.\\
            \textbf{SUS 2} & I found the system unnecessarily complex.\\
            \textbf{SUS 3} & I thought the system was easy to use.\\
            \textbf{SUS 4} & I think that I would need the support of a technical person to be able to use this system.\\
            \textbf{SUS 5} & I found the various functions in this system were well integrated.\\
            \textbf{SUS 6} & I thought there was too much inconsistency in this system.\\
            \textbf{SUS 7} & I would imagine that most people would learn to use this system very quickly.\\
            \textbf{SUS 8} & I found the system very cumbersome to use.\\
            \textbf{SUS 9} & I felt very confident using the system.\\
            \textbf{SUS 10} & I needed to learn a lot of things before I could get going with this system.\\
            \bottomrule
        \end{tabular}
        \label{tab:SUS items}
    \end{table}

\section{SUS Questionnaire Responses: AUXILIO}\label{app:SUS ratings_auxilio}
    \begin{table}[htbp]
        \centering
        \label{tab:SUS ratings_auxilio}
        \resizebox{\textwidth}{!}{%
        \begin{tabular}{ccccccccccccc}
            \toprule
            \textbf{Respondent} & \textbf{SUS1} & \textbf{SUS2} & \textbf{SUS3} & \textbf{SUS4} & \textbf{SUS5} & \textbf{SUS6} & \textbf{SUS7} & \textbf{SUS8} & \textbf{SUS9} & \textbf{SUS10} & \textbf{SUS Score} & \textbf{Grade} \\ 
            \midrule
            \textbf{R1} & 5	& 1 & 4	& 2	& 5	& 1	& 5	& 2	& 5	& 1 & 92.50 & A+\\
            \textbf{R2} & 3	& 1	& 5	& 2	& 5	& 2	& 5	& 2	& 4	& 1 & 85.00 & A+\\
            \textbf{R3} & 5	& 2	& 4	& 1	& 5	& 2	& 4	& 2	& 4	& 1 & 85.00 & A+\\
            \textbf{R4} & 3	& 1	& 5	& 2	& 5	& 1	& 4	& 2	& 5	& 1 & 87.50 & A+\\
            \textbf{R5} & 1	& 2	& 2	& 2	& 4	& 2	& 3	& 4	& 2	& 1 & 52.50 & D\\
            \textbf{R6} & 3	& 2	& 3	& 2	& 4	& 2	& 5	& 2	& 3	& 2 & 70.00 & C \\
            \textbf{R7} & 2	& 2	& 5	& 2	& 4	& 2	& 5	& 1	& 5	& 1 & 82.50 & A \\
            \textbf{R8} & 2	& 1	& 4	& 2	& 3	& 2	& 3	& 2	& 4	& 1 & 70.00 & C \\
            \textbf{R9} & 4	& 2	& 4	& 5	& 5	& 1	& 4	& 2	& 4	& 2 & 72.50 & C+ \\
            \textbf{R10} & 4 & 1 & 5 & 1 & 4 & 1 & 5 & 1 & 4 & 2 & 90.00 & A+ \\
            \midrule
            \textbf{Mean} & 3.20 & 1.50 & 4.10 & 2.10 & 4.40 & 1.60 & 4.30 & 2.00 & 4.00 & 1.30 & 78.75 & \multirow{2}{*}{B+} \\
            \textbf{SD} & 1.32 & 0.53 & 0.99 & 1.10 & 0.70 & 0.52 & 0.82 & 0.82 & 0.94 & 0.48 & 12.32 &  \\ \bottomrule
        \end{tabular}%
        }
    \end{table}

\section{SUS Questionnaire Responses: Smyle Mouse}\label{app:SUS ratings_smyle}
    \begin{table}[htbp]
        \centering
        \label{tab:SUS ratings_smyle}
        \resizebox{\textwidth}{!}{%
        \begin{tabular}{ccccccccccccc}
            \toprule
            \textbf{Respondent} & \textbf{SUS1} & \textbf{SUS2} & \textbf{SUS3} & \textbf{SUS4} & \textbf{SUS5} & \textbf{SUS6} & \textbf{SUS7} & \textbf{SUS8} & \textbf{SUS9} & \textbf{SUS10} & \textbf{SUS Score} & \textbf{Grade} \\ 
            \midrule
            \textbf{R1} & 5 & 1 & 5 & 1 & 5 & 1 & 5 & 1 & 2 & 1 & 92.50 & A+\\
            \textbf{R2} & 2 & 1 & 4 & 2 & 5 & 1 & 5 & 2 & 4 & 1 & 82.50 & A\\
            \textbf{R3} & 1 & 2 & 2 & 2 & 2 & 3 & 1 & 3 & 2 & 2 & 40.00 & F\\
            \textbf{R4} & 1 & 2 & 2 & 3 & 1 & 5 & 4 & 5 & 2 & 2 & 32.50 & F\\
            \textbf{R5} & 1 & 2 & 2 & 2 & 4 & 5 & 3 & 3 & 2 & 2 & 45.00 & F\\
            \textbf{R6} & 2 & 4 & 1 & 2 & 2 & 4 & 2 & 4 & 3 & 3 & 32.50 & F\\
            \textbf{R7} & 1 & 1 & 4 & 2 & 5 & 3 & 4 & 3 & 3 & 2 & 65.00 & C\\
            \textbf{R8} & 1 & 2 & 3 & 2 & 2 & 5 & 2 & 5 & 1 & 3 & 30.00 & F\\
            \textbf{R9} & 2 & 4 & 2 & 4 & 2 & 3 & 3 & 5 & 2 & 3 & 30.00 & F\\
            \textbf{R10} & 2 & 3 & 3 & 2 & 2 & 5 & 3 & 4 & 2 & 3 & 37.5 & F\\
            \midrule
            \textbf{Mean} & 1.80 & 2.20 & 2.80 & 2.20 & 3.00 & 3.50 & 3.20 & 3.50 & 2.30 & 2.20 & 48.75 & \multirow{2}{*}{F} \\
            \textbf{SD} & 1.23 & 1.14 & 1.23 & 0.79 & 1.56 & 1.58 & 1.32 & 1.35 & 0..82 & 0.79 & 23.01 &  \\ \bottomrule
        \end{tabular}%
        }
    \end{table}

\section{Themes and Codes Summary: Auxilio} \label{app:themescodes_auxilio}
    \begin{table}[H]
        \centering
        \label{tab:themesauxilio}
    
        \begin{tabular}{l c}
            \toprule
            \textbf{Themes}/Codes  &  Frequency\\
            \midrule
            \textbf{Ease of Use and Deterministic Behavior} & \textbf{29}\\
            Easy Click Mechanisms (LMB+RMB) & 9\\
            Persistent Calibration & 6\\
            Easy Device Calibration & 5\\
            Easy Pointer Movement & 5\\
            Deterministic Device Behavior & 4\\
            \midrule
            \textbf{Difficult Precise and Static Pointing} & \textbf{21}\\
            Oscillating Pointer Movement & 8\\
            Difficult Small Target Clicking & 7\\
            Difficult Static Targeting & 5\\
            Unrecognized Clicks & 1\\
            \midrule
            \textbf{Ergonomics, Cognitive Load, and Comfort} & \textbf{12}\\
            Less Head Movement & 3\\
            No/Minimal Strain from Head Movements & 3\\
            Comfortable Design & 3\\
            No/Minimal Muscle Fatigue & 2\\
            Lower Cognitive Load & 1\\
            \midrule
            \textbf{High Responsiveness} & \textbf{7}\\
            Easy Dynamic Target Tracking & 3\\
            Easy Large/Non-Small Target Hitting & 2\\
            Easy Cursor Control at Extreme Angles & 1\\
            Faster Task Completion & 1\\
            \midrule
            \textbf{Adaptability} & \textbf{6}\\
            Positive/Easy Adaptation & 4\\
            Initial Learning Phase & 2\\
            \midrule
            \textbf{Discomforts} & \textbf{5}\\
            Visual Obstruction from Device & 3\\
            Problem for Special Cases/Disabilities & 1\\
            Uncomfortable Design & 1\\
            \midrule
            \textbf{Additional Features} & \textbf{1}\\
            Extra Controls and Features & 1\\
            \bottomrule
        \end{tabular}
    \end{table}

\section{Themes and Codes Summary: Smyle Mouse} \label{app:themescodes_smyle}
    \begin{table}[H]
        \centering
        \label{tab:themessmyle}
    
        \begin{tabular}{l c}
            \toprule
            \textbf{Themes}/Codes  &  Frequency\\
            \midrule
            \textbf{Performance and Calibration Issues} & \textbf{51}\\
            Unrecognized Clicks & 18\\
            Frequent Recalibration & 13\\
            Cursor Drift Over Time & 8\\
            Click Mechanism Drift Over Time & 6\\
            Non-deterministic Device Behavior & 2\\
            Discrete Pointer Movement & 2\\
            Difficult Dynamic Target Tracking & 2\\
            \midrule
            \textbf{Ergonomic Issues, Cognitive Load, and Physical Discomforts} & \textbf{23}\\
            Neck Strain from Head Movements & 8\\
            More Head Movement & 4\\
            Visual Discomfort & 4\\
            Discomfort from Prolonged Use & 3\\
            Higher Cognitive Load & 2\\
            Jaw Muscle Fatigue & 2\\
            \midrule
            \textbf{Design and Learnability Issues} & \textbf{14}\\
            Difficult Cursor Control at Extreme Angles & 6\\
            Difficult Click Mechanisms (LMB+RMB) & 4\\
            Difficult Device Calibration & 2\\
            Constrained Performance & 1\\
            Difficult Pointer Movement & 1\\
            \midrule
            \textbf{Stable Static Pointing} & \textbf{5}\\
            Stable Pointer Movement & 3\\
            Easy Static Targeting & 2\\
            \midrule
            \textbf{Ease of Use} & \textbf{3}\\
            Easy Click Mechanisms (LMB+RMB) & 1\\
            Easy Device Calibration & 1\\
            Easy Small Target Clicking & 1\\
            \bottomrule
        \end{tabular}
    \end{table}

\section{Cursor Movement Heatmaps of Auxilio and Smyle Mouse} \label{app:heatmaps}
    \newlength{\heatmapwidth}%
        \setlength{\heatmapwidth}{0.42\textwidth}%
        \begin{longtable}[c]{|c|c|c|}
        
            \hline
            ID & Auxilio & Smyle Mouse\\
            \hline
            \endfirsthead
    
            \hline
            \multicolumn{3}{|c|}{Continuation of the Heatmap Table}\\
            \hline
            ID & Auxilio & Smyle Mouse\\
            \hline
            \endhead
    
            \hline
            \endfoot
    
            \hline
            \endlastfoot
    
            1 & 
            \raisebox{-.5\height}{\includegraphics[width=\heatmapwidth]{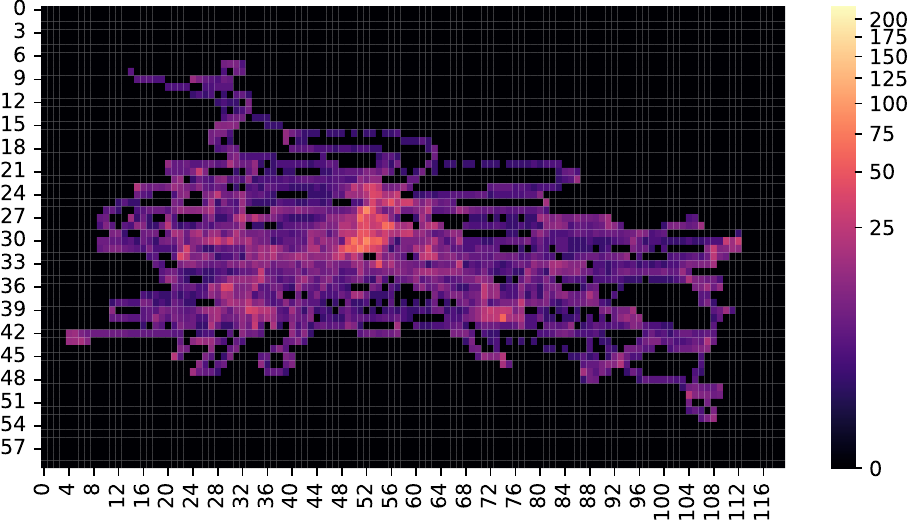}} &
            \raisebox{-.5\height}{\includegraphics[width=\heatmapwidth]{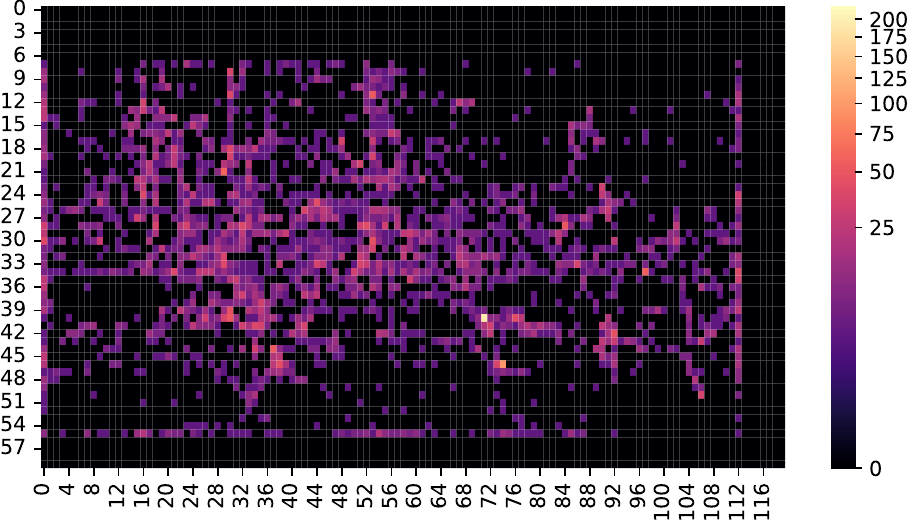}}\\ \hline
            2 & 
            \raisebox{-.5\height}{\includegraphics[width=\heatmapwidth]{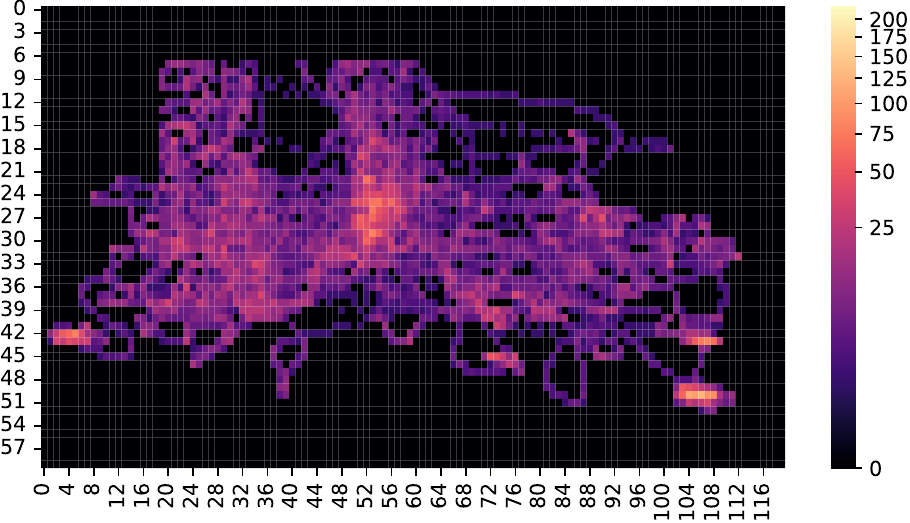}} &
            \raisebox{-.5\height}{\includegraphics[width=\heatmapwidth]{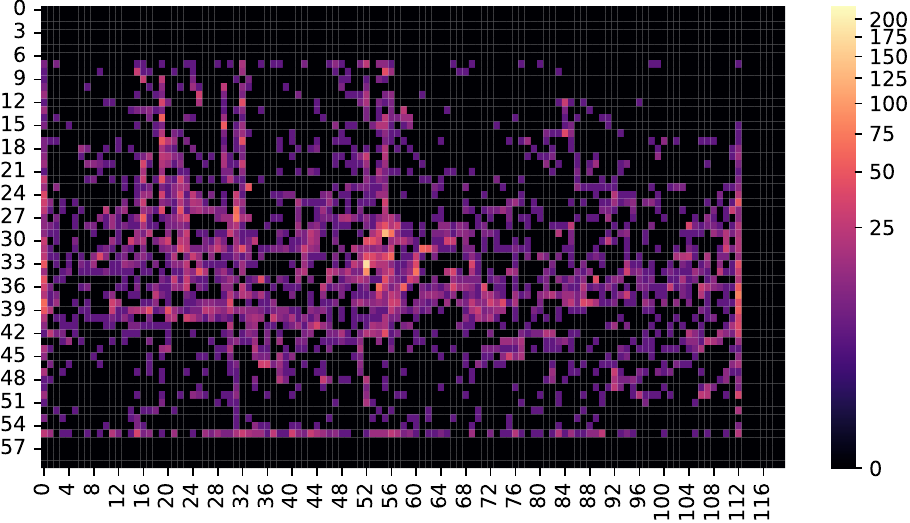}}\\ \hline
            3 & 
            \raisebox{-.5\height}{\includegraphics[width=\heatmapwidth]{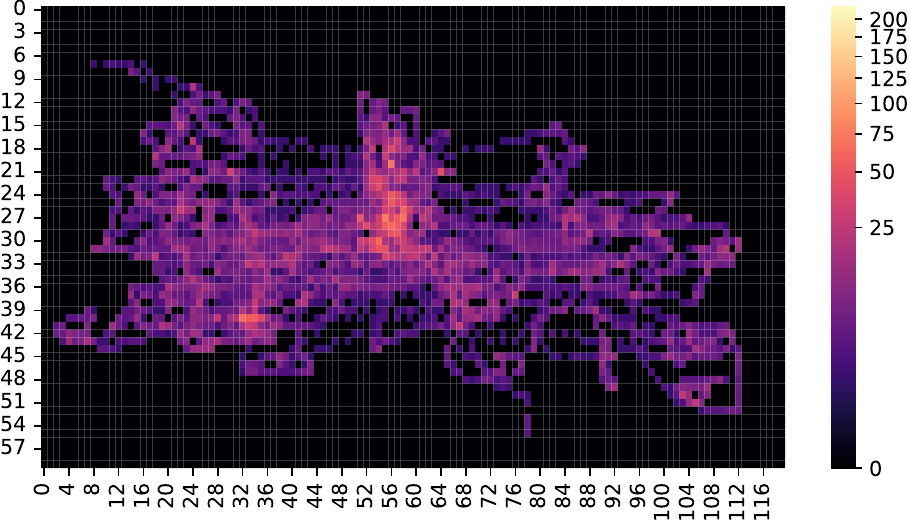}} &
            \raisebox{-.5\height}{\includegraphics[width=\heatmapwidth]{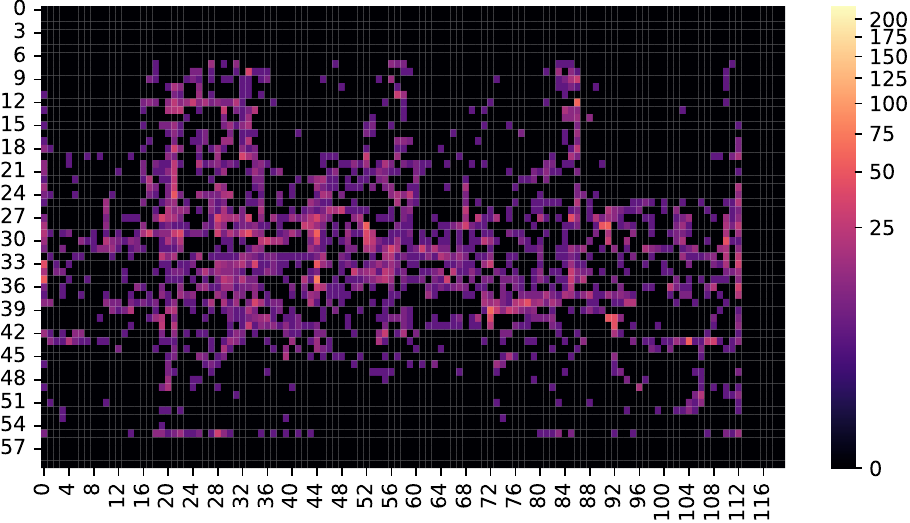}}\\ \hline
            4 & 
            \raisebox{-.5\height}{\includegraphics[width=\heatmapwidth]{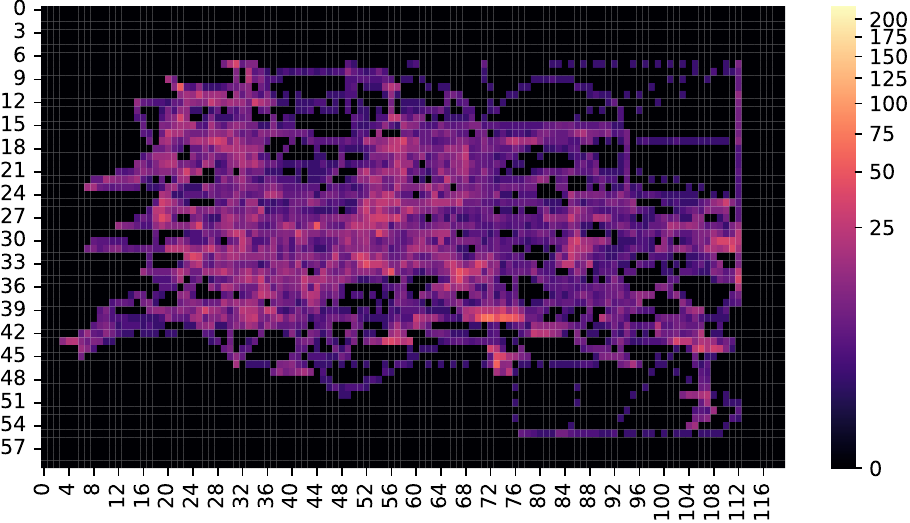}} &
            \raisebox{-.5\height}{\includegraphics[width=\heatmapwidth]{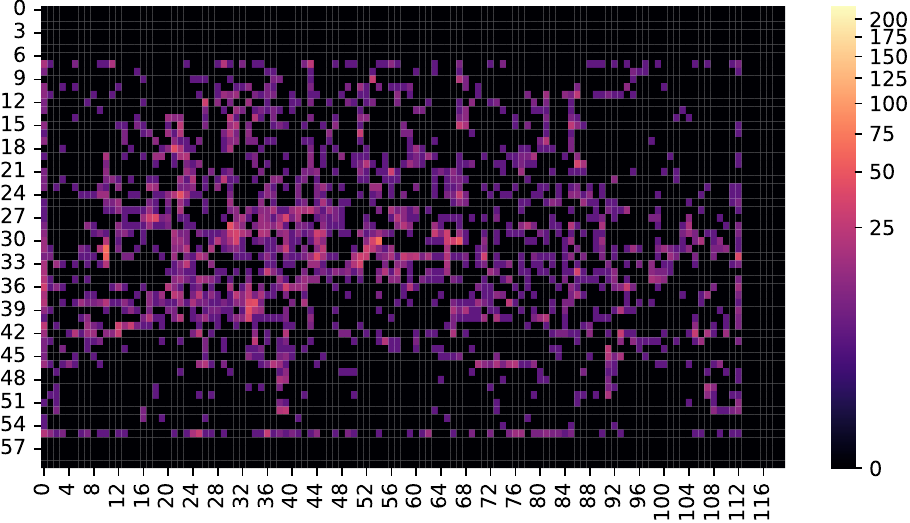}}\\ \hline
            5 & 
            \raisebox{-.5\height}{\includegraphics[width=\heatmapwidth]{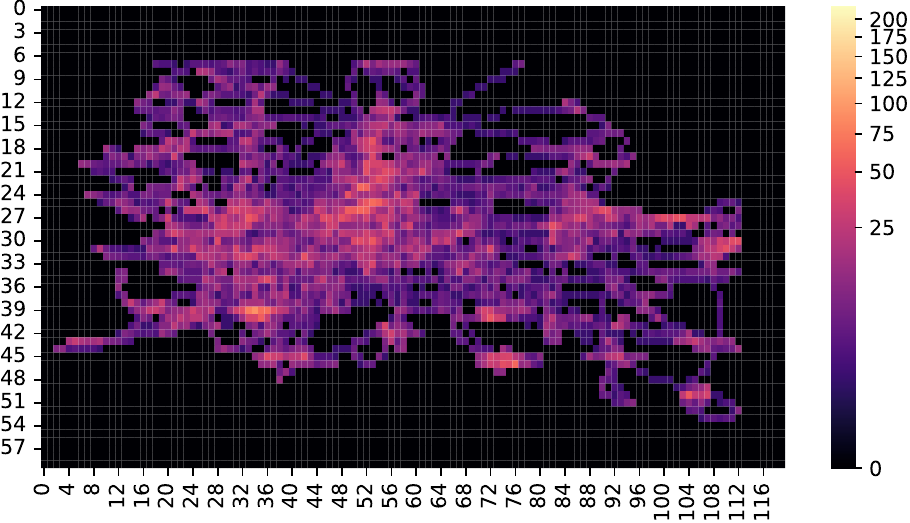}} &
            \raisebox{-.5\height}{\includegraphics[width=\heatmapwidth]{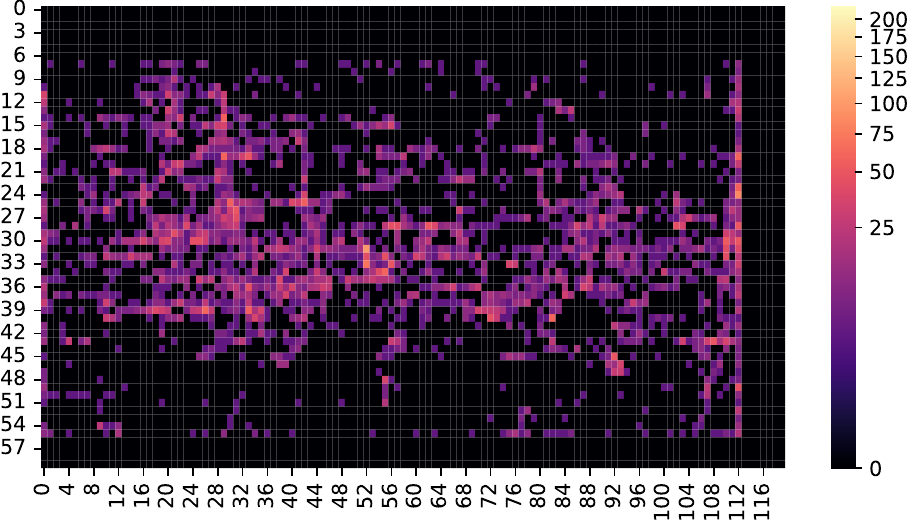}}\\ \hline
            6 & 
            \raisebox{-.5\height}{\includegraphics[width=\heatmapwidth]{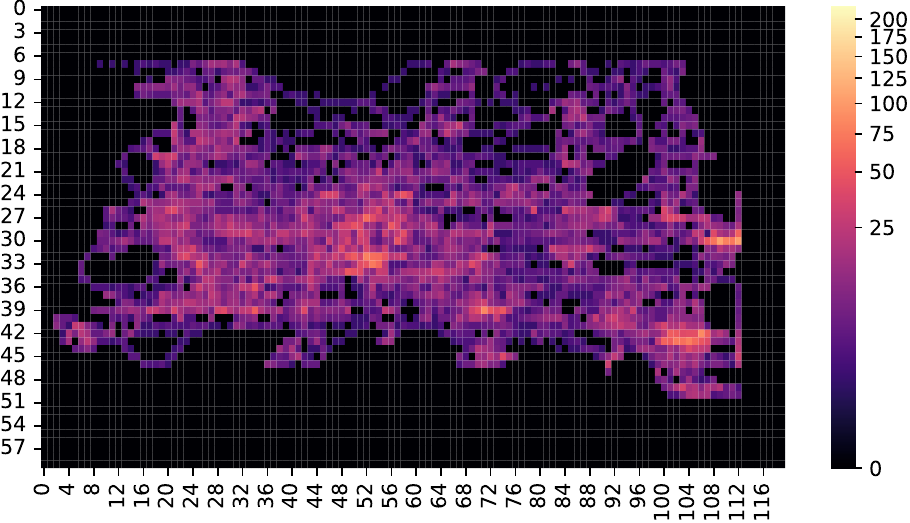}} &
            \raisebox{-.5\height}{\includegraphics[width=\heatmapwidth]{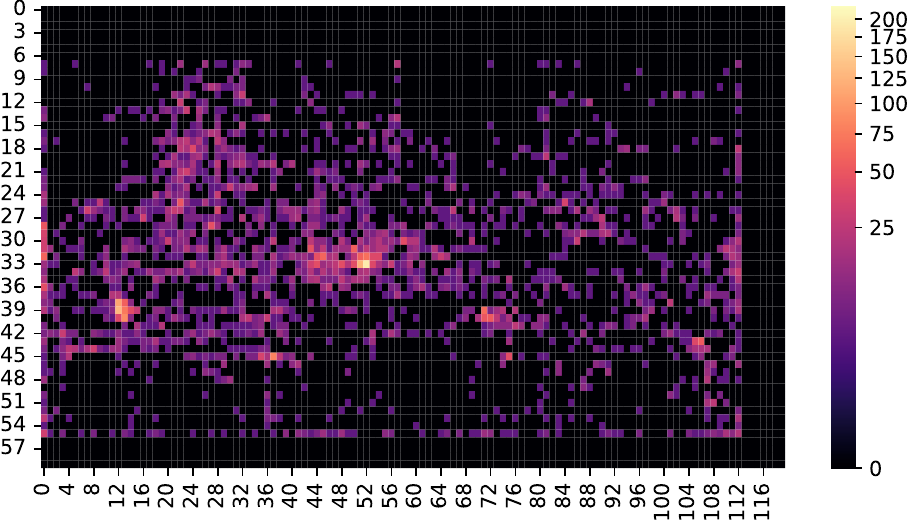}}\\ \hline
            7 & 
            \raisebox{-.5\height}{\includegraphics[width=\heatmapwidth]{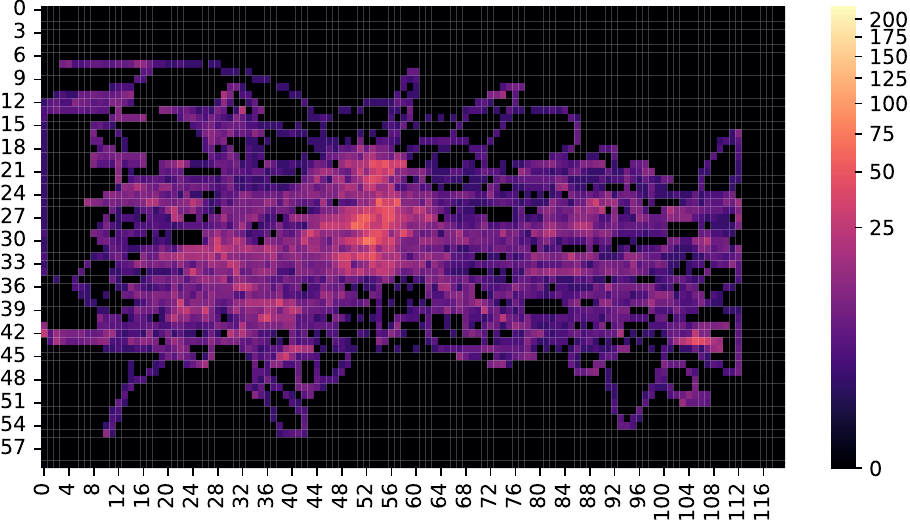}} &
            \raisebox{-.5\height}{\includegraphics[width=\heatmapwidth]{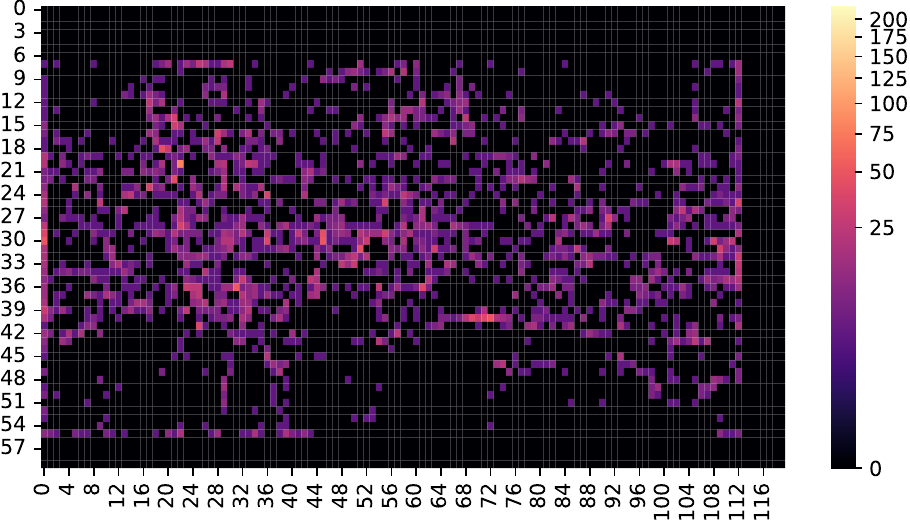}}\\ \hline
            8 & 
            \raisebox{-.5\height}{\includegraphics[width=\heatmapwidth]{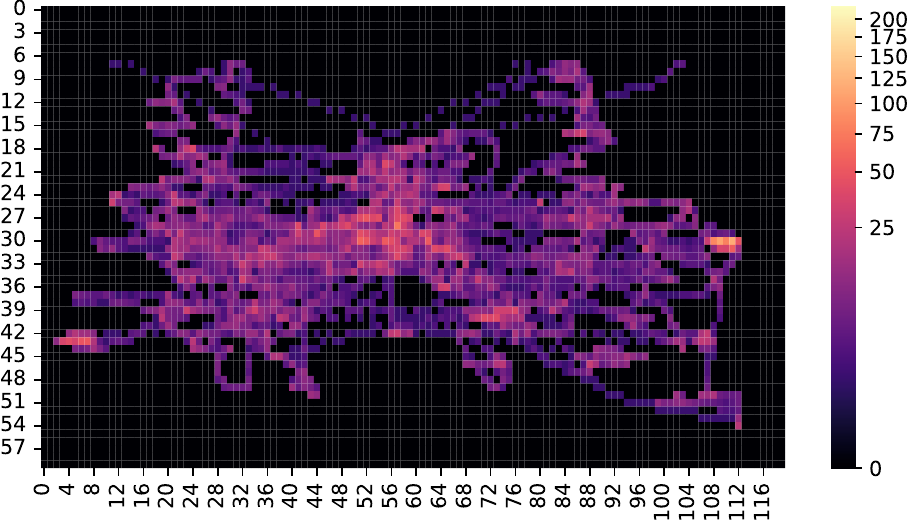}} &
            \raisebox{-.5\height}{\includegraphics[width=\heatmapwidth]{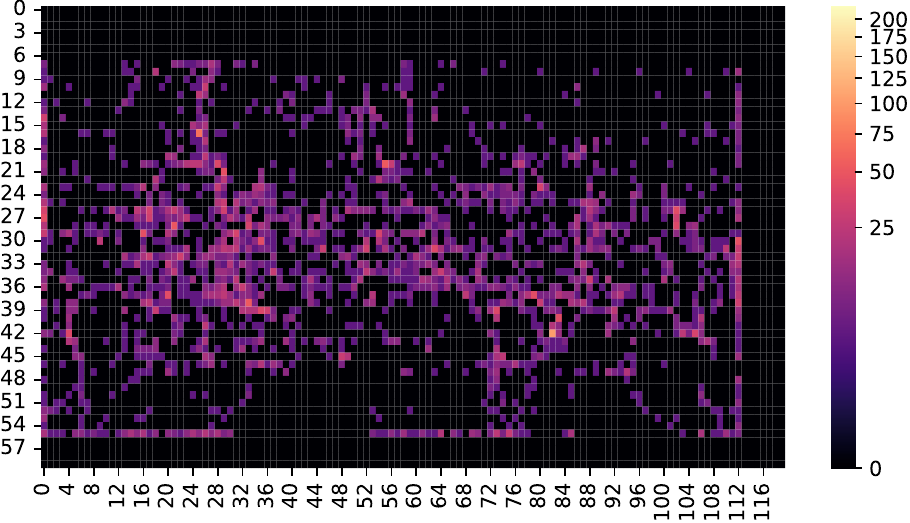}}\\ \hline
            9 & 
            \raisebox{-.5\height}{\includegraphics[width=\heatmapwidth]{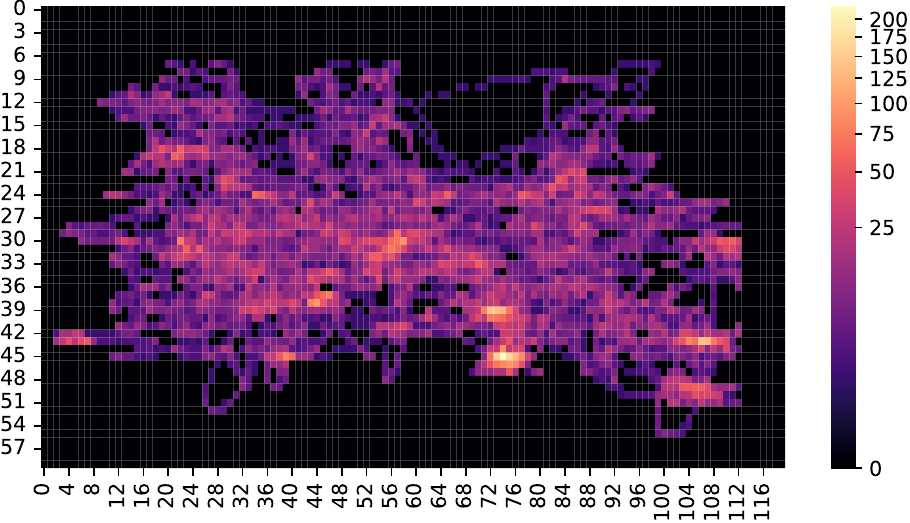}} &
            \raisebox{-.5\height}{\includegraphics[width=\heatmapwidth]{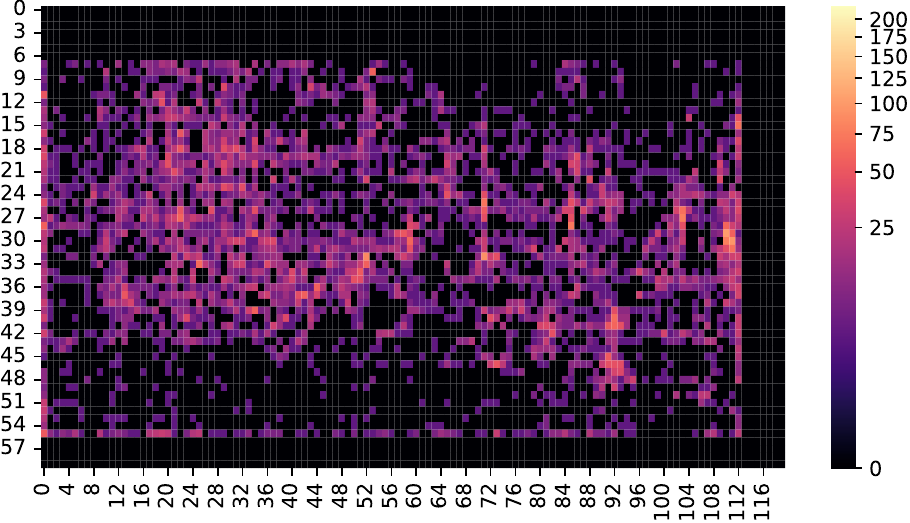}}\\ \hline
            10 & 
            \raisebox{-.5\height}{\includegraphics[width=\heatmapwidth]{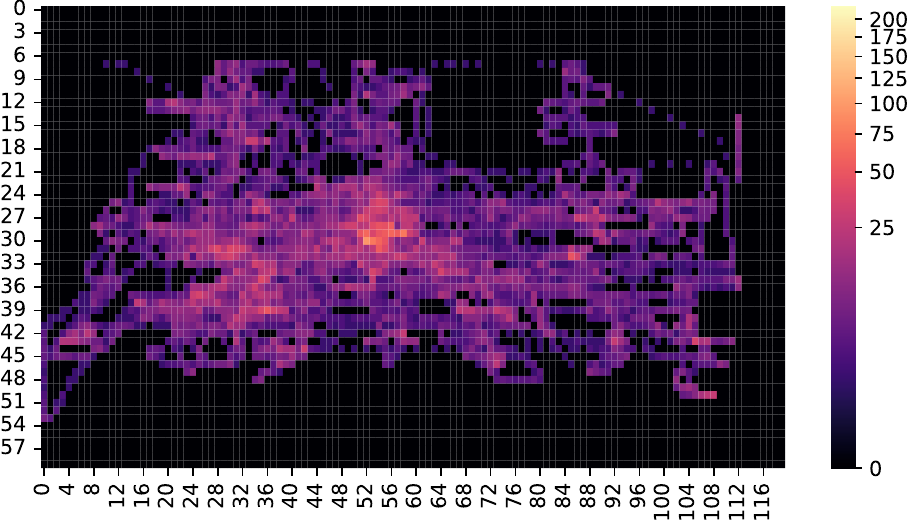}} &
            \raisebox{-.5\height}{\includegraphics[width=\heatmapwidth]{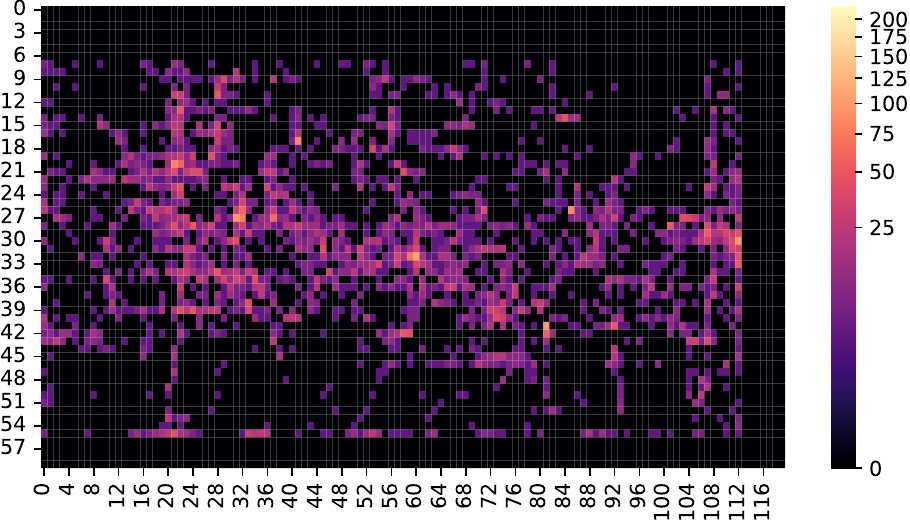}}\\ \hline
        \end{longtable}

\section{Consent Form} \label{app:consent}
    \begin{center}
    \Huge \textbf{Informed Consent Form --- Auxilio}
\end{center}
\section*{Study Title}
Point--and--Click Experiment to Compare \emph{Auxilio} with other input device(s)

\section*{Principal Investigators}
\begin{itemize}
    \item \censor{Mohammad Ridwan Kabir, Assistant Professor, Department of Computer Science and Engineering (CSE)}
    \item \censor{Mohammad Ishrak Abedin, Lecturer, Department of Computer Science and Engineering (CSE)}
\end{itemize}

\section*{Institution}
\censor{Islamic University of Technology (IUT), Board Bazar, Gazipur-1704, Bangladesh}

\section*{Contact Information}
\begin{itemize}
    \item \censor{Mohammad Ridwan Kabir} --- Mail: \censor{\href{mailto:ridwan.kabir@queensu.ca}{ridwan.kabir@queensu.ca}}
    \item \censor{Mohammad Ishrak Abedin} --- Mail: \censor{\href{mailto:ishrakabedin@iut-dhaka.edu}{ishrakabedin@iut-dhaka.edu}}
\end{itemize}

\section*{Purpose of the Study}
You are invited to participate in a research study aimed at comparing \emph{Auxilio} against different input device(s) through a point-and-click experiment. As part of this study, we will also collect data on skin conductance, temperature, and heart rate to assess potential stress levels during the experiment.

\section*{Procedures}
If you agree to participate in this study, you will be asked to perform a series of tasks using different input devices. Your age and sex will be recorded for understanding the participant demographics. During the experiment, we will record your performance data as well as your skin conductance, temperature, and heart rate using non-invasive sensors.

\section*{Duration}
The experiment will take approximately 35--50 minutes to complete.

\section*{Voluntary Participation}
Your participation in this study is entirely voluntary. You may withdraw from the study at any time without any penalty or loss of benefits to which you are otherwise entitled.

\section*{Confidentiality}
All data collected during this study will be kept anonymous. Your personal identity will not be linked to the data in any way, and only aggregated data will be used for analysis and publication.

\section*{Potential Risks and Discomforts}
The risks associated with this study are minimal and are comparable to those encountered in everyday use of computer input devices. The sensors used to monitor your skin conductance, temperature, and heart rate are non-invasive and should not cause any discomfort.

\section*{Potential Benefits}
While there may be no direct benefits to you, your participation will contribute to a better understanding of the effectiveness and stress implications of different input devices.

\section*{Compensation}
Each participant will be paid \censor{500 BDT} for their participation in this study.

\section*{Questions}
If you have any questions about the study, you may contact any of the principal investigators at the contact information provided above. 

\section*{Consent}
By signing below, you acknowledge that you have read and understood the information provided above, and you agree to participate in this study. You also understand that you can withdraw your consent at any time without penalty.

\vspace{2cm}

\noindent
Participant's Name: \hrulefill

\vspace{1cm}

\noindent
Participant's Signature: \hrulefill

\vspace{1cm}

\noindent
Date: \hrulefill

\end{document}